\documentclass[11pt,reqno]{amsproc}
\usepackage[german,english]{babel}  
\usepackage[latin1]{inputenc} 
\usepackage{graphicx, color}
\usepackage{amssymb}
\numberwithin{equation}{section}
\newtheorem{theorem}{Theorem}[section]
\newtheorem{proposition}[theorem]{Proposition}
\newtheorem{lemma}[theorem]{Lemma}
\newtheorem{corollary}[theorem]{Corollary}
\newtheorem{definition}[theorem]{Definition}

\newtheorem{remark}[theorem]{Remark}

\def\tr{\mathop{\textnormal{tr}}}
\newcommand{\loc}{\text{\rm{loc}}}
\newcommand{\unif}{\text{\rm{unif}}}
\newcommand{\supp}{\text{\rm{supp}}}
\newcommand{\R}{\mathbb{R}}

\newcommand{\Z}{\mathbb{Z}}

\renewcommand{\P}{\mathbb{P}}
\newcommand{\E}{\mathbb{E}}
\newcommand{\N}{\mathbb{N}}
\newcommand{\1}{\mathbb{1}}

\newcommand{\Rd}{\mathbb{R}^{d}}
\newcommand{\LL}{\Lambda_L}

\newcommand{\no}[1]{|\,#1|}
\newcommand{\Poxty}{\P_{0,x}^{t,y}}
\newcommand{\Eoxty}{\E_{0,x}^{t,y}}

\def\<{\langle}
\def\>{\rangle}

\def\p0{\psi_0}                              

\def\1Ll{{1 \over {\vert \Lambda_L \vert}}}

\def\j1d{{(1+\vert j \vert )}^{-(d+2)}}
\def\x1a{{(1+ \vert x\vert )}^{-\alpha}}
\def\2j1a{{(1+ \vert j\vert )}^{-\alpha}}

\begin{document}

\title[IDS for Random Schr\"odinger Operators]
{The Integrated Density of States\\ for Random Schr\"odinger Operators}

\author{Werner Kirsch}
\author[Bernd Metzger]{Bernd Metzger$^*$}
\thanks{$^*$Supported by the Deutsche Forschungsgemeinschaft (DFG), SFB--TR 12}
\address{Institut f\"ur Mathematik,
Ruhr-Universit\"at Bochum, D-44780 Bochum, Germany}
\email{werner.kirsch@ruhr-uni-bochum.de}
\email{bernd.metzger@ruhr-uni-bochum.de}

\dedicatory{Dedicated to Barry Simon on the occasion of his
60th birthday}

\vspace{5mm}


\begin{abstract} We survey some aspects of the theory of the
integrated density of states (IDS) of random Schr\"odinger
operators. The first part motivates the problem and introduces the
relevant models as well as quantities of interest. The proof of
the existence of this interesting quantity, the IDS, is discussed
in the second section. One central topic of this survey is the
asymptotic behavior of the integrated density of states at the
boundary of the spectrum. In particular, we are interested in Lifshitz tails and the
occurrence of a classical and a quantum regime. In the last
section we discuss regularity properties of the IDS. Our emphasis is
on the discussion of fundamental problems and central
ideas to handle them. Finally, we discuss further developments and
problems of current research.
\end{abstract}

\maketitle

\clearpage

\setcounter{tocdepth}{2} \tableofcontents

\clearpage

\section{Introduction\label{sec:intro}}

\subsection{Physical Background: Models and Notation}
The time evolution of a quantum mechanical state $\psi$ is
obtained from the time dependent Schr\"odinger equation
\begin{equation}\label{eq:3.0}
i\,\frac{\partial}{\partial t}\,\psi \;= \;H\,\psi .
\end{equation}
Another important differential equation is the heat or diffusion
equation
\begin{equation}\label{eq:3.01}
 \,\frac{\partial}{\partial t}\,\psi \;= - \;H\,\psi .
\end{equation}
In order to investigate either of these equations, it is extremely
useful to know as much as possible about the operator $H$ and its
spectrum. In general the Schr\"odinger operator $H$ is of the form
\begin{displaymath}
H=H_0+V.
\end{displaymath}
The  free operator $H_0$   represents the kinetic energy of the
particle. In the absence of magnetic fields, it is given by the
Laplacian
\begin{displaymath}
H_0=- \,\Delta=- \,\sum_{\nu=1}^{d}\,\frac{\partial^2}{{\partial
x_\nu}^2}.
\end{displaymath}
The potential $V$ encoding the forces $F(x) =-\nabla{V}(x)$ is
acting as a multiplication operator in the Hilbert space
$L^2(\R^d)$.\\

Occasionally, we will replace $H_0$ with the operator $H_0(B)$ which
contains a homogeneous magnetic field. For $d=2$  $H_0(B)$, $B>0$,
is given by
\begin{eqnarray*}
H_0(B)=\left(i\frac{\partial}{\partial
x_1}-\frac{1}{2}Bx_2\right)^2+\left(i\frac{\partial}{\partial
x_2}+\frac{1}{2}Bx_1\right)^2.
\end{eqnarray*}
In contrast to $H_0(B=0)$, the spectrum of $H_0(B)$ for $B>0$ is a
countable set $\sigma(H_0(B))=\{(2n+1)B; n\in\N\}$. The energies
$E_n=(2n+1)$ are eigenvalues of $H_0(B)$ of infinite multiplicity,
called the Landau levels.

\subsection{Random Potentials\label{sec:RandPot}}
First-order physical modeling often assumes an ideal
background, e.g., a homogeneous material without any impurities. For
example, in ideal crystals the atoms or nuclei are supposed to be
distributed on a periodic lattice (say the lattice $\Z^d$) in a
completely regular way. We assume that a particle (electron) at the
point $x\in\R^d$ feels a potential of the form $q\, f(x-i) $ due to
an atom (or ion or nucleus) located at the point $i\in\Z^d$. We call
the function $f$ the \emph{single site potential}. The coupling
constant $q$ represents the charge of the particle at the lattice
point $i$. So, in a regular crystal, the particle is exposed to a
total potential

\begin{eqnarray}\label{eq:4.0}
V(x)=\sum_{i\in\Z^d}\;q\,f(x-i).
\end{eqnarray}
The potential $V$ in (\ref{eq:4.0}) is periodic with respect to the
lattice $\Z^d$, i.e., $V(x-i)=V(x)$ for all $x\in\R^d$ and
$i\in\Z^d$. The mathematical theory of Schr\"odinger operators with
periodic potentials is well developed (see, e.g., \cite{Eastham,
Kuchment, ReedSimon4}). It is based on a thorough analysis of the
symmetry properties of periodic operators. The spectrum of periodic
Schr\"odinger operators has a band structure and the spectrum is
purely absolutely continuous, i.e.,
\begin{eqnarray}
\sigma(H)=\bigcup_{n=0}^\infty\;
[a_n, b_n] \qquad \textnormal{with }~ a_n<b_n\le a_{n+1},\\
\sigma_{\text{sing}}(H) = \emptyset, \qquad\qquad \sigma(H)=\sigma_{\text{ac}}(H).
\end{eqnarray}
The real world is not ideal. Solids occur in nature in various
forms. Sometimes they are (almost) totally ordered, sometimes they
are more or less completely disordered. For a mathematical modelling
of disordered solids two ingredients are essential: the spatial
homogeneity in the mean and the disappearance of long range
correlations. In full generality these properties are studied in the
theory of ergodic operators. This class of operators contains, e.g.,
 quasicrystals and can also be related to random matrices.
In this publication we will be concerned with \emph{random}
Schr\"odinger
operators.

The most popular  and best understood model of a disordered solid is
the alloy-type Schr\"odinger operator. It models a mixture of
different atoms located at lattice positions. The type of atom at
the lattice point $i$ is assumed to be random. These particles are
represented by randomly distributed coupling constants $q_i$
encoding the different charges. The total potential is given by

\begin{equation}\label{def:Vomega}
 V_\omega(x)=\sum_{i\in\Z^d}\;q_i(\omega)\,f(x-i).
 \end{equation}
When talking about the alloy-type model, we will mean
(\ref{def:Vomega}) with the following assumptions:
\begin{enumerate}
\item \label{ass:fbound}The single site potential $f$  is bounded,
non-negative and strictly positive on an open set. \item
\label{ass:fdecay}$f$ satisfies $f(x)\le
C\,\big(1+\no{x})^{-(d+\varepsilon)}$ for some $C$ and
$\varepsilon>0$. \item The random variables $q_i$ are independent
and identically distributed random variables on a probability space
$(\Omega,\mathcal{F},\P)$. \item The common probability distribution
of the $q_i$ is denoted by $P_0$. Its support $\supp\,P_0$ is
compact and contains at least two points.
\end{enumerate}
Assumption (\ref{ass:fbound}) can be considerably relaxed. For
example, for most of the following results one may allow local
singularities for $f$. By assuming (\ref{ass:fbound}) we avoid
technical difficulties which may obscure the main argument. More
details on weaker conditions can be found in the papers cited.
Assumption (\ref{ass:fdecay}) ensures that the sum in
(\ref{def:Vomega}) is convergent. The compactness of $\supp\,P_0$ is
convenient but not always necessary and in some especially marked
situations we consider also unbounded single site distributions.
However, for many of our results we need that the $q_i$ (and hence
$\supp\,P_0$) are bounded \emph{below}. The physical model suggests
that $\supp\,P_0$ consists of finitely many points only (the
possible charges). However, for many mathematical results it is
necessary (or at least convenient) to suppose that $P_0$ has a
(bounded) density $g$,
i.e., $P_0=g(\lambda)\,d\lambda$.

One might argue that such an assumption is acceptable as a purely
technical one. On the other hand one could argue the problem is not
understood  as long as it is impossible to handle the physically
relevant case of finitely many
values.

A simplified version of the alloy-type  Schr\"odinger operator above
is the (discrete) Anderson model. Here the Hilbert space is the
sequence space $\ell^2(\Z^d)$ instead of $L^2(\R^d)$ and the free
operator $H_0$ is replaced by the discrete analogue of the
Laplacian. This is the finite-difference operator
\begin{equation} (h_0\,u)(n) = -\sum_{|m-n|=1}\,(u(m)-u(n)).  \label{def:H0}
\end{equation}
Above we set $|n|=\sum_{i=1}^{d} |n_i|$ on $\Z^d$. The
potential  is a multiplication operator $V=V_\omega$ on
$\ell^2(\Z^d)$ with $V_\omega(i)$ independent, identically
distributed, and the total Hamiltonian is defined by
$h=h_0+V_\omega$. We will call this setting the ``discrete case'' in
contrast to Schr\"odinger operators on $L^2(\R^d)$ which we refer
to as the ``continuous case.''

The most frequently used and most important approach to model
\emph{amorphous material} like glass or rubber defines the random
locations of the atoms by a Poisson random measure. This random
point measure can be characterized by the number $n_A =
\mu_\omega(A)$ of random points in the set $A$. We assume that the
random variables $n_A$ and $n_B$ are independent for disjoint
(measurable) sets $A$, $B$ and   $\P\,\big(n_A =
k\big)=\frac{{|A|}^{k}}{k!}\;e^{-|A|}$ ($|A|$ denotes the Lebesgue
measure of $A$). With this notation we may write the Poisson
potential for an amorphous solid as

\begin{equation}\label{mod:Poi}
V_\omega(x)=\int_{\R^d}\,q\,f(x-\eta)\;d\,\mu_\omega(\eta) .
\end{equation}
\\[2mm]
To model thin disordered layers, we also consider random potentials
which are concentrated along a hypersurface in $\R^d$ (or $\Z^d$).
For example, we are going to consider ``surface'' alloy potentials. To
define such a potential let us write $\R^d=\R^{d_1}\times\R^{d_2}$
then
\begin{eqnarray*}
V_\omega(x_1,x_2)=\sum_{i_1\in\Z^{d_1}}q_{i_1}(\omega)f(x_1-i_1,x_2)
\end{eqnarray*}
is a random potential which is concentrated along the hypersurface
$\R^{d_1}$ in $\R^d$. In addition, there may be a random or periodic
background potential on $\R^d$.

Most of the theorems we are going to discuss
can be proved for rather general single site potentials $f$ and
probability distributions $P_0$ of the $q_i$. For example, most of
the time we can allow some local singularities of $f$. To simplify
the following discussion, we will assume in this paper the conditions
defined in the context of the alloy-type potential.

The above random operators are examples of ``ergodic operators.''
This class of operators includes not only most random operators but
also periodic and almost periodic operators. Most of the results of
Section~\ref{sec:ids} and part of Section~\ref{sec:regular} can be
shown for general ergodic operators. We refer to
\cite{PF,CL,CFKS,Lifshitz3,Stollmann2} and \cite{Kirsch3} for a
discussion of this general context.

\subsection{The Concept of the Integrated Density of States}
The (integrated) density of states is a concept of
fundamental importance in condensed matter physics. It measures the
``number of
energy levels per unit volume'' near (resp.\ below) a given energy.

Typical systems arising in solid state physics have periodic or
ergodic potentials. Consequently, the spectrum of the corresponding
Hamiltonian is not discrete. Therefore, we cannot just count the
eigenvalues below $E$ or within an interval $[E_1,E_2]$. On the
other hand, the number of electrons in such a system, which extends
to infinity,  ought to be infinite. For these two reasons, the Pauli
exclusion principle does not make immediate sense. (How do we
distribute infinitely many electrons on a continuum of spectral
energies?)

However, there may be a chance to make sense out of the Pauli
principle by first restricting the system to a finite volume
$\Lambda$. Inside $\Lambda$ there should be only finitely many
electrons, in fact, we may   assume that the number of electrons in
a given $\Lambda$ is proportional to the
volume $|\Lambda|$ of this set.

If $P$ is a finite-dimensional orthogonal projection, then $\tr(P)$
is the dimension of its range. If $P_{(-\infty,E]}$ is the spectral
projection of a random Schr\"odinger operator (which as a rule has
infinite-dimensional range) and if $\Lambda_L$ is a cube of side
length $L$ around the origin, then we may call
$\tr(\chi_{\Lambda_L}P_{(-\infty,E]})$ the restriction of
$P_{(-\infty,E]}$ to the cube $\LL$. Above $\chi_A$ is the
characteristic function of the set $A$. Thus, we may try to define
the integrated density of states as
\begin{equation}\label{IDS0}
 N(E)\;=\;\lim_{L\rightarrow
\infty }\;
\frac{1}{|\LL|}\;\tr\,\big(\chi_{\LL}P_{(-\infty,E]}\big).
\end{equation}
Of course, we have to prove that the limit in (\ref{IDS0}) does
exist and is not trivial. We will deal with these questions in
Section 2.

There is another way to define the integrated density of states
which turns out to be equivalent to (\ref{IDS0}):   We
restrict the operator $H_\omega$ to the Hilbert space
$L^2(\Lambda)$. To obtain a self adjoint operator we have to impose
boundary conditions at $\partial\Lambda$, e.g., Dirichlet or
Neumann boundary conditions. We call the corresponding operators
$H_\Lambda^D$ and $H_\Lambda^N$, respectively. These operators have
compact resolvents, i.e., their spectra are purely discrete. We
denote by
\begin{equation}
E_1(H_\Lambda^D)\le E_2(H_\Lambda^D)\le E_3(H_\Lambda^D)\ldots
\end{equation}
the eigenvalues of $H_\Lambda^D$ (and analogously for $H_\Lambda^N$)
in increasing order, where eigenvalues are repeated according to
their multiplicity. The eigenvalue counting function of an operator
$A$ with purely discrete spectrum is defined by
\begin{equation}
    N(A,E)=\#\{\,n\,|\,E_n(A)\le E\}\;=\;\tr\big(P_{(-\infty,
    E]}(A)\big).
\end{equation}
Analogously to (\ref{IDS0}), we can therefore define
\begin{align}\label{IDSD}
    N^D(E)&= \lim_{L\rightarrow
\infty }\; \frac{1}{|\LL|}\;N(H_{\LL}^D,E)\notag\\
&= \lim_{L\rightarrow \infty }\;
\frac{1}{|\LL|}\;\tr\big(P_{(-\infty,E]}(H_{\LL}^D)\big)
\end{align}
and similarly for Neumann boundary conditions,
\begin{align}\label{IDSN}
    N^N(E)&= \lim_{L\rightarrow
\infty }\; \frac{1}{|\LL|}\;N(H_{\LL}^N,E)\notag\\
&= \lim_{L\rightarrow \infty }\;
\frac{1}{|\LL|}\;\tr\big(P_{(-\infty,E]}(H_{\LL}^N)\big).
\end{align}
This procedure to define the integrated density of states  makes
sense only if $N$, $N^D$ and $N^N$ all \emph{exist}
 and \emph{agree}.

This is, indeed, the case. We will see in the sequel that each of
these definitions has its own technical advantage. The integrated
density of states $N$ is basic for studying the physical (in
particular the thermodynamical) properties of disordered systems.
>>From a mathematical point of view, the properties of $N$ are
interesting in their own respect. Moreover, properties of $N$
constitute an essential input to prove localization properties of the system.

It is the aim of this review to discuss some of the problems and
results connected with the integrated density of states. In Section~2
we sketch the proof of the existence of the integrated density
of states and discuss some fundamental questions concerning the
probabilistic and the functional analytic approach. In Section~3
we study the behavior of the integrated density of states at
the boundary of the spectrum. In the last section we discuss some
basic ideas concerning the regularity of the integrated density
of states.

\medskip
\noindent\textbf{Acknowledgements.} It is a pleasure to thank many
colleagues for fruitful collaborations and stimulating discussions
on the subject. There are too many to name them here. We also would
like to stress the fact that the selection of topics within our
subject and the way of presenting them is due to our very personal
preferences. We have most certainly left out important topics and
works. This is to be blamed on our ignorance and the limitation of
time and space.

We would like to thank Jessica Langner and Riccardo
Catalano for their skillful typing and careful proofreading of the
manuscript.

\goodbreak
\section{The Density of States Measure: Existence\label{sec:ids}}

\subsection{Introduction and Historical Remarks}
The first existence proofs for the integrated density of
states go back at least to Pastur. See \cite{Pastur3}
for an early review of the subject.

There are a couple of methods to prove the existence of the
integrated density of states. One of them, invented and used by
Pastur, is based on the Laplace transform of the integrated density
of states and of its approximants. For this method, one proves the
convergence of the Laplace transform and uses the fact that
convergence of the Laplace transform implies the vague convergence
of the underlying measures.

To prove the convergence of the Laplace transforms, it is useful to
express the Laplace transform of the finite-volume quantities using
the Feynman--Kac-representation of the Schr\"odinger semigroup
$e^{-tH}$. Feynman--Kac and Laplace transform methods are also used
to prove the equivalence of the definitions of the integrated
density of states (\ref{IDS0}) and (\ref{IDSD}), (\ref{IDSN}) with
either Neumann or Dirichlet (or more general) boundary conditions
(see, e.g., Pastur \cite{Pastur3} or \cite{KM1,DK,Kirsch3}).\\
The definition of the integrated density of states via (\ref{IDS0})
was used by Avron and Simon in the context of almost periodic
potentials \cite{AS2}. They also proved that the spectrum of the
operator coincides with the growth points of the integrated density
of states. In Section~\ref{ex:ids} we will follow this approach
to prove the existence of the integrated density of states.

One of the virtues of the definition of the integrated density of
states via boundary conditions (\ref{IDSD}) and (\ref{IDSN}) is the
fact that they allow lower and upper bounds of $N$  ``for free.'' In
fact, one way to examine the   behavior of $N$ at the bottom (or top
$=\infty$) of the spectrum is based on this approach. We will
discuss this approach in Section~\ref{sec:ND} and the estimates
based on it in Section~3.

As a rule, quantitative estimates on the effect of introducing
boundary conditions are hard to obtain. For example, if one
investigates the behavior of $N$ at internal spectral edges it seems
extremely difficult to control the perturbation of eigenvalues due
to boundary conditions. Klopp \cite{Klopp3} proposed an
approximation of the random potential by periodic ones with growing
period. This way we lose monotonicity which is at the heart of the
Neumann--Dirichlet approach. Instead one can prove that the
approximation is exponentially fast thus gaining good estimates of
the remainder.

Finally, we would like to mention that one can also define the
integrated density of states via Krein's spectral shift function.
This reasoning is well known in scattering theory (see, e.g.,
\cite{Birman,Yafaev}). In connection with random Schr\"odinger
operators, the spectral shift function was first used by Simon
\cite{Simon8} to investigate spectral averaging. Kostrykin and
Schrader \cite{KS1} applied this technique to prove the existence of
the integrated density of states and the density of surface states.
This method turns out to be useful also to investigate regularity
properties of the integrated density of states \cite{KS2}.

The results of this section are true not only for the specific
random potentials discussed in Section~\ref{sec:intro}, but rather
for general ergodic operators. In fact, the proofs carry over to
this general setting in most cases. We refer to the survey
\cite{Kirsch3} for details.

\subsection{The Existence of the Integrated Density of States\label{ex:ids}}
In this section we prove  the existence of the integrated
density of states as defined in (\ref{IDS0}). To do so, we need
little more than Birkhoff's ergodic theorem (see, e.g.,
\cite{Lamperti}). Below, as in the rest of this paper, we denote by
$\E$ the expectation with respect to the probability measure $\P$.

\begin{proposition}\label{P1}
 If $\varphi$ is a bounded measurable function of compact support, then
 \begin{equation}\label{eq:b1}
    \lim_{L\rightarrow\infty}\frac{1}{|\Lambda_L|}\;\tr\,(\varphi(H_\omega)\chi_{\Lambda_L})=
    \E\,\big(\tr\,(\chi_{\Lambda_1}\varphi(H_\omega)\chi_{\Lambda_1})\big)
 \end{equation}
 for $\P$-almost all $\omega$.
\end{proposition}
\begin{proof}
Define $\xi_i=\tr\,(\varphi(H_\omega)\chi_{\Lambda_1(i)})$. $\xi_i$
is an ergodic sequence of random variables. Hence, Birkhoff's ergodic
theorem implies that
\begin{equation*}
\frac{1}{|\Lambda_L|}\;\tr\,(\varphi(H_\omega)\chi_{\Lambda_L})\;=\;\frac{1}{|\Lambda_L|}\;\sum_{i\in\LL}\;\xi_i
\end{equation*}
converges to its expectation value.
\end{proof}

The right-hand side of (\ref{eq:b1}) as well as
$|\Lambda_L|^{-1}
\,\tr\left(\,\varphi(H_\omega)\,\chi_{\Lambda_L}\right)$ are
positive linear functionals on the bounded, continuous functions.
They define positive measures $\nu$ and $\nu_L$ by
\begin{displaymath}
    \int{\varphi(\lambda)d\nu(\lambda)}=\E(\,\big(\tr\,(\chi_{\Lambda_1}\varphi(H_\omega)\chi_{\Lambda_1})\big))
\end{displaymath}
and
\begin{displaymath}
\int_\R \,\varphi(\lambda)\;d\nu_L(\lambda)
=\frac{1}{|\Lambda_L|}\tr(\varphi(H_\omega)\chi_{\Lambda_L}).
\end{displaymath}
Equation (\ref{eq:b1}) suggests that the measures $\nu_L$ might
converge to the limit measure $\nu$ as $L\to\infty$ in the sense of
vague convergence of measures. The problem is (\ref{eq:b1}) holds
only for fixed $\varphi$ on a set $\Omega_\varphi$ of full
probability; respectively  (\ref{eq:b1}) holds for \emph{all}
$\varphi$ for $\omega\in\bigcap_{\varphi}\Omega_\varphi$. However,
this is an \emph{uncountable} intersection of sets of probability
one. The problem is solved by  approximating   $C_0(\R)$ by a
countable, dense subset.
\begin{theorem}\label{T1}
 The measures $\nu_L$ converge vaguely to the measure $\nu$ $\P$-almost surely, i.e., there is a set $\Omega_0$ of
 probability one, such that
 \begin{equation}\label{eq:b2}
  \int\varphi(\lambda)d\nu_L(\lambda)\rightarrow \int\varphi(\lambda)d\nu(\lambda)
 \end{equation}
 for all $\varphi\in C_0(\R)$, the set of continuous functions with compact support, and all $\omega\in\Omega_0$.
 \end{theorem}
\noindent
\begin{definition}
The non-random probability measure $\nu$ is called the density of
states measure. The distribution function $N$ of  $\nu$, defined by
\begin{displaymath}
    N(E)=\nu(]-\infty,E]) ,
\end{displaymath}
is known as the integrated density of states.
\end{definition}

Using Theorem \ref{T1} it is not hard to see:

\begin{proposition}[\cite{AS2}]
 $\supp(\nu)=\Sigma\,[=\sigma(H_\omega)\quad a.s.]$.
\end{proposition}
\vspace{5mm}
\subsection{Existence via Dirichlet--Neumann-Bracketing\label{sec:ND}}
Our first approach to define the density of states measure was based on the additivity of
$\tr(\varphi(H_\omega)\chi_{\Lambda_L})$ and the ergodic theorem by
Birkhoff. This very naturally fits in the concept of self-averaged
quantities from physics.

However, for some part of the further analysis, an alternative
approach---the Dirichlet--Neumann bracketing---is more suitable. Let
$(H_\omega)_{\Lambda}^N$ and $(H_\omega)_{\Lambda}^D$ be the
restrictions of $H_{ \omega} $ to $L^2(\Lambda)$ with Neumann and
Dirichlet boundary conditions. See, e.g., \cite{ReedSimon4} for an
appropriate definition of these boundary conditions via quadratic
forms.  Furthermore, we define (for $X = N \textnormal{ or } D,
\textnormal{ and } E\in \R)$
\begin{eqnarray}\label{def:NX}
  N_{\Lambda}^X(E)\,:=\, N(\,(H_\omega)_{\Lambda}^{X}, E) &=&
   \tr(\chi_{(-\infty,E]}({H_\omega}_\Lambda^{X})) .
\end{eqnarray}
Our aim is to consider the limits
\begin{eqnarray}
 \label{def:limNX}
  N^{X} (E) &=&
  \lim_{L\rightarrow\infty}\;\; \frac{1}{\,|\Lambda_L|\,}\;
  N_{\LL}^X(E).
\end{eqnarray}
The quantities $N_\Lambda^D$ and $N_\Lambda^N$ as defined in
(\ref{def:NX}) are distribution functions of point measures
$\nu_\Lambda^D$ and $\nu_\Lambda^N$ concentrated in the eigenvalues
of $H_\Lambda^D$ and $H_\Lambda^N$, i.e.,
\begin{equation}
N_\Lambda^X(E)~=~\nu_\Lambda^X\big((-\infty,E]\big).
\end{equation}
The convergence in (\ref{def:limNX}) is meant as the vague
convergence of the corresponding measures or, what is the same, as
the pointwise convergence of the distribution function
$\frac{1}{|\Lambda|}\,N_\Lambda^X$ at all continuity points of the
limit.

Let us first look at ${1\over \vert \Lambda\vert}\,N_{\Lambda}^D
(E)$.   The random field $N_{\Lambda}^D $ is {\it not} additive in
$\Lambda$, so that we \emph{cannot} use Birkhoff's ergodic theorem.
However, $N_{\Lambda}^D $ is   \emph{superadditive}, in the sense
that $N_{\Lambda}^D \,(E)\ge\;N_{\Lambda_1}^D(E)
\;+\;N_{\Lambda_2}^D (E) $ whenever $\Lambda = \Lambda_1 \cup
\Lambda_2$ with $(\Lambda_1)° \cap (\Lambda_2)° = \emptyset$. ($M°$
denotes the interior of the set $M$.)  Similarly, $N_{\Lambda}^N$ is
{\it subadditive}, i.e., $- N^N$ is superadditive.

\begin{theorem} $N_{\Lambda}^{D}$ is superadditive and $N_{\Lambda}^{N}$
is subadditive. More precisely, if $\Lambda =\Lambda_{1}\cup
\Lambda_{2}$ and $(\Lambda_1)° \cap (\Lambda_2)° = \emptyset$ then
$$N_{\Lambda_{1}}^{D}(E) +N_{\Lambda_{2}}^{D}(E)\le N_{\Lambda}^{D}(E)\le N_{\Lambda}^{N}(E)\le
N_{\Lambda_{1}}^{N}(E)+N_{\Lambda_{2}}^{N}(E).$$
\end{theorem}

Fortunately, there are sub- and superadditive versions of the
ergodic theorem, going back at least to Kingman \cite{King}. The
situation  here is ideal for the superadditive ergodic theorem by
Akcoglu and Krengel \cite{AK}. Indeed, one can prove that (for fixed
$E$) the processes $N_{\Lambda}^{D}(E)$ and $N_{\Lambda}^{N}(E)$ are
superadditive and subadditive random fields in the sense of
\cite{AK} respectively (see \cite{KM1,Krengel}). This yields the
following result.

\begin{theorem}[\cite{KM1}]\label{th:exist}
The limits
$$
\bar{N}^{D}(E)=\lim_{L\to\infty}~~{1\over \vert
\Lambda_{L}\vert}~N(H_{\omega~{\Lambda}}^{D},E)
$$
and
$$\bar{N}^{N}(E)=\lim_{L\to\infty}~~{1\over \vert
\Lambda_{L}\vert}~N(H_{\omega~{\Lambda}}^{N},E)
$$
exist $P$-almost surely.  Moreover,
\begin{align*}
\bar{N}^{D}(E) &= \sup_{L}~~\frac{1}{\vert
\Lambda_{L}\vert}\;\;\E\,\big(N(H_{\omega~{\Lambda_{L}}}^{D},E)\big) \\
\bar{N}^{N}(E)&= \inf_{L}~~\frac{1}{\vert
\Lambda_{L}\vert}\;\;\E\,\big(N(H_{\omega~{\Lambda_{L}}}^{N},E)\big).
\end{align*}
\end{theorem}

The functions $\bar{N}^X$ are increasing functions.
However, it is not clear whether they are right continuous. To
obtain \emph{distribution functions}, we  define $N^X$ by making the
$\bar{N}^X$ right continuous
\begin{align}
N^D(E)&=  \inf_{E'>E}\;\bar{N}^D(E')\\
N^N(E) &= \inf_{E'>E}\;\bar{N}^N(E').
\end{align}
Note that $\bar{N}^X$ and $N^X$ disagree at most on a
countable set. Since $N^D$ are $N^N$ are distribution functions, they
define measures by
\begin{align}
\nu^D\big((a, b]\big) &= N^D(b)-N^D(a)\\
\nu^N\big((a, b]\big) &= N^N(b)-N^N(a).
\end{align}

>>From Theorem \ref{th:exist} we obtain the following
corollary, which we will use to investigate the asymptotic behavior
of the integrated density of states (e.g., for small $E$).

\begin{corollary} \label{DNB} For any $\Lambda$,
$${1\over \vert \Lambda
\vert}~\E\,\big(N({H_{\omega}}_{\Lambda_{L}}^{D}, E~)\big) \le~
\bar{N}^D(E)~\le~\bar{N}^N(E)~\le~ {1\over\vert \Lambda\vert}~
\E\,(N({H_{\omega}}_{\Lambda_{L}}^{N}, E)).$$
\end{corollary}
\noindent Our physical intuition would lead to the hope that
$\,N(E)\,=\,N^{D}(E)=\,N^{N}(E)$ since, after all, the introduction
of boundary conditions was a mathematical artifact that should not
play any role for the final physically meaningful quantity. This is,
in fact, true under fairly weak conditions (see \cite{Pastur3, KM1,
DK} and references given there).

\begin{theorem}\label{th:uniq}
The distribution functions $N(E)$, $N^D(E)$ and $N^N(E)$ agree.
\end{theorem}

Theorem \ref{th:uniq} follows from Theorem \ref{FKrep} in
the next section. An alternative proof for the Anderson model can be
found in the review \cite{Kirsch5}. Theorem \ref{th:uniq} implies a
fortiori that the quantities $\frac{1}{\,|\Lambda_L|\,}\;
N_{\LL}^D(E)$ and $\frac{1}{\,|\Lambda_L|\,}\; N_{\LL}^D(E)$
converge to the same limit, except for a countable set of energies $E$.
The exceptional points, if any, are the discontinuity
points of $N$. We will discuss continuity (and, more generally,
regularity) properties of $N$ in Section 4.

\subsection{A Feynman--Kac Representation for $N$\label{sec:FK}}
In this section we will consider the Laplace transform
of the integrated density of states (both $N(E)$ and $N^X(E)$).
The Laplace transform of a measure $\nu$ with distribution function
$F$ is defined by
\begin{equation}\label{def:LT}
\widetilde{\nu}(t)=\widetilde{F}(t):=\int\,e^{-\lambda
t}\;d\nu(\lambda)\,=\;\int\,e^{-\lambda t}\;dF(\lambda).
\end{equation}

There is a very useful representation of the Laplace
transform of $N$ (and of $N^X$) via the Feynman--Kac formula. Using
this representation one can show that $N$ and $N^X$ are, indeed,
the same quantities. Moreover, the Feynman--Kac-representation of $N$
is very useful to compute the asymptotic behavior of $N$ for
small or large energies.

The key ingredients of the representation formula for
$\widetilde{N}(t)$ are the Brownian motion, the Brownian bridge and
the Feynman--Kac-formula. For material about these concepts, we refer
to Reed--Simon  \cite{ReedSimon2,ReedSimon4}   and Simon
\cite{Simon5}.

By $\Poxty$ we denote the measure underlying a Brownian bridge
starting in the point $x\in\R^d$ at time $0$ and ending at time $t$
in the point $y$. $\Eoxty $ denotes integration over $\Poxty $. A
Brownian bridge is a Brownian motion $b$ conditioned on $b(t)=y$.
Note that $\Poxty$ is not a probability measure. $\Poxty$ has total
mass $p(t,x,y)$ where $p$ denotes the probability kernel of the
Brownian motion.

\begin{theorem}[Feynman--Kac formula]\label{th:FK} If $V \in L^P_{\loc,\, \unif}\,
(\Rd)$ for $p =2$ if  $d \le 3$, $p > d/2$ if $d\ge 3 $, then
$e^{-tH}$ has a jointly continuous integral kernel given by
\begin{align}\label{FK}
e^{-tH}(x,y) &=  \Eoxty (e^{-{\int_0^t}V(b(s))\, ds} )\\
\nonumber &= \int {e^{-{\int_0^t}V(b(s))\, ds}} d\Poxty (b).
\end{align}
The integration here is over paths $ b(.)\in C ([0,t])$.
\end{theorem}

We remind the reader that we always assume bounded
potentials so that the conditions in Theorem \ref{th:FK} are
satisfied. In the context of the density of states, we are interested
in a Feynman--Kac formula for Hamiltonians on {\sl bounded} domains.
Let us denote by $\Omega^t_\Lambda $ the set of all paths staying
inside $\Lambda$ up to time $t$, i.e.,
$$
\Omega_\Lambda^t = \left\{ b \in C ([0,t]: \vert b(s) \in \Lambda
\hbox{ for all } 0 \le s \le t \right\}.
$$
Then $e^{-t{H_\Lambda^D}} $ is simply given by restricting the
integration in (\ref{FK}) to the set $\Omega_\Lambda^t $, e.g.,
$$
e^{-t{H_\Lambda^D}}(x,y)   = \Eoxty
\big(e^{-{\int_0^{t}\,V\left(b(s)\right)\, ds}}
{\chi}_{{\Omega_\Lambda^t}}\big).
$$
A proof can be found in Simon \cite{Simon5} and Aizenman--Simon
\cite{Aizenman}. There is also a Feynman--Kac formula for Neumann
boundary conditions (see \cite{KM1} and references given there).

Now we are able to state the probabilistic representation of the
density of states measure in terms of Brownian motion.

\begin{theorem}\label{FKrep}
The Laplace transforms of $N(E)$, $N^D(E)$ and $N^N(E)$ agree and
are given by
\begin{equation}\label{fkn}
 \widetilde{N}(t)=\widetilde{N}^D(t) = \widetilde{N}^N(t) = \E\times
\E_{0,0}^{t,0}(e^{-\int_{0}^{t}V_{\omega}(b(s))ds}).
\end{equation}
\end{theorem}

For a detailed proof see, e.g., \cite{Kirsch3} and
references there.  The first step in the proof is to check that the
right-hand side of (\ref{fkn}) is finite for all $t\ge0$.
Interchanging the expectation values with respect to random
potential and the Brownian motion, this
follows from Jensen's inequality.

The second step of the proof is to compare (\ref{fkn}) and the
Laplace transforms of the approximating density of states measure
$\nu_L$ and $\nu^D_L$. To prove the theorem one has to estimate the
hitting probability of the boundary of $\Lambda$ for a Brownian
motion starting and ending far away from the boundary. Using
standard facts of Brownian motion, this tends to $0$ in the
limit $|\Lambda| \rightarrow \infty$.

Once we know that the Laplace transforms of $N$, $N^D$ and $N^N$
agree, it follows from the uniqueness of the Laplace transform that
$N$, $N^D$ and $N^N$ agree themselves (see, e.g., \cite{Feller}).

\subsection{The Density of Surface States}
We would like to define a density of states measure for
surface potentials as well. Suppose we have a surface
potential of the form
\begin{eqnarray*}
V^s_\omega(x_1,x_2)=\sum_{i_1\in\Z^{d_1}}q_{i_1}(\omega)f(x_1-i_1,x_2)
\end{eqnarray*}
where, as above, $x\in\R^d$ is written as $x=(x_1,x_2)$ with
$x_1\in\R^{d_1}, x_2\in\R^{d_2}$.
In addition to the surface potential, there may be a random or
periodic potential $V^b(x)$, which we call the ``bulk'' potential. The
bulk potential should be stationary and ergodic with respect to
shifts $T_j$ parallel to the surface. Stationarity perpendicular to
the surface is not required in the following.

This allows  ``interfaces'' in the following sense: Let $d_1=d-1$, so
the surface has codimension one. Thus it forms the interface between
the upper half space $V_+=\{x; x_2>0\}$ and the lower half space
$V_-$. The bulk potential $V^b$ may then be defined by
$V^b(x)=V_1(x)$ for $x_2\ge 0$ and $=V_2(x)$ for $x_2<0$. Here $V_1$
and $V_2$ are random or periodic potentials on $\R^d$. We set
$H^b=H_0+V^b$, which we call the ``bulk operator'' and
$H_\omega=H^b+V^s_\omega$.

We could try to define a density of states measure in the same way
as in (\ref{eq:b1}), i.e., look at
\begin{equation}\label{def:doss0}
 \lim_{L\rightarrow\infty}\frac{1}{L^d}\tr(\varphi(H_\omega)
    \chi_{\Lambda_L}).
\end{equation}
It is not hard to see that this limit exists and equals
\begin{equation}
   \E\,\big(\tr\,(\chi_{\Lambda_1}\varphi(H^b)\chi_{\Lambda_1})\big).
\end{equation}
In other words, (\ref{def:doss0}) gives the density of
states measure for the bulk operator. After all, this is not really
surprising. The normalization with the volume term $L^d$ is
obviously destroying any influence of the surface potential.

So it sounds reasonable to choose a surface term like $L^{d_1}$ as
normalization and to consider
\begin{equation}\label{def:doss1}
 \lim_{L\rightarrow\infty}\frac{1}{L^{d_1}}\tr(\varphi(H_\omega)
    \chi_{\Lambda_L}).
\end{equation}
However, Definition (\ref{def:doss1}) gives a finite result only
when
$\supp\,\varphi\cap\sigma(H^b)=\emptyset$.

To define the density of surface states also inside the spectrum of
the bulk operator, we therefore set
\begin{equation}\label{def:doss}
 \nu_s(\varphi)= \lim_{L\rightarrow\infty}\frac{1}{L^{d_1}}\tr\Big(\big(\varphi(H_\omega)-\varphi(H^b)\big)
    \chi_{\Lambda_L}\Big).
\end{equation}
Of course, it is not obvious at all that the limit (\ref{def:doss})
exists. In \cite{EKSS1, EKSS2} the authors proved that the limit
exists for functions $\varphi\in C_0^3(\R)$. Hence the density of
surface states is defined as a distribution. The order of this
distribution is at most $3$. Observe that, in contrast to the
density of states, the limit in (\ref{def:doss}) is not necessarily
positive for positive $\varphi$ due to the subtraction term. In
fact, in the discrete case it is not hard to see that the total integral
of the density of states, i.e.,  $\nu_s(\mathbf{1})$, is zero.
Therefore, we cannot conclude that the density of surface states is
a (positive) measure.

Kostrykin and Schrader \cite{KS1,KS2} proved that the density of
surface states distribution is actually the derivative of a
measurable locally integrable function. They do not prove that this
function is of bounded variation, thus leaving the possibility that
$\nu_s$ is \emph{not} given by a measure. See also the papers
\cite{Chahrour1, Chahrour2} by Chahrour for regularity properties of
the density of surface states on the lattice.

Outside the spectrum of $H^b$, the distribution $\nu_s$ is positive,
so that the density of surface states \emph{is} a measure
\emph{there}. In \cite{KiWa2} it was proven that below the ``bulk''
spectrum $\sigma(H^b)$ the density of surface states can also be
defined by using (Neumann or Dirichlet) boundary conditions. We
expect this to be wrong inside the bulk spectrum.

\section{Lifshitz Tails}
\subsection{The Problem}\label{intro}
For a periodic potential $V$ the integrated density of
states $N(E)$ behaves near the bottom $E_0$ of the spectrum
$\sigma(H_0+V)$ like
\begin{equation}\label{os}
    N(E)\sim C(E-E_{0})^{d/2}.
\end{equation}
This can be shown by explicit calculation for $V\equiv 0$ and was
proved for general periodic potentials in \cite{KSJFA}.

On the basis of physical arguments Lifshitz \cite{Lifshitz1,
Lifshitz2} predicted a completely different behavior for disordered
systems, namely,
\begin{equation}\label{ds}
    N(E)\sim C_{1}\ e^{-C_{2}(E-E_{0})^{-d/2}}
\end{equation}
as $E\searrow E_{0} > -\infty$. This behavior of $N(E)$ is called
{\it Lifshitz behavior} or {\it Lifshitz tails}. The reason for this
peculiar behavior is a collective phenomenon. To simplify the
following heuristic argument, let us assume that $V_\omega\ge 0$ and
$E_0=0$. To find an eigenvalue smaller than $E$, the potential
$V_\omega$ has to be small on a rather large region in space. In
fact, to have an eigenvalue at small $E>0$, the uncertainty
principle (i.e., the kinetic energy) forces the potential to be
smaller than $E$ on a set whose volume is of the order $E^{-d/2}$.
That $V_\omega$ is small on a large set is a typical ``large
deviations event'' which is very rare---in fact, its probability is
exponentially small in terms of the volume of the set, i.e., its
probability is of the order
\begin{equation}\label{est:exp}
    e^{-C_2\,E^{-d/2}}
\end{equation}
which is precisely the behavior (\ref{ds}) predicted by Lifshitz. It
is the aim of this section to discuss the Lifshitz behavior
(\ref{ds}) of the integrated density of states as well as its
extensions and limitations.

The first proof of Lifshitz behavior (for the Poisson model
(\ref{mod:Poi})) was given by Donsker and Varadhan \cite{DV}. They
estimated the Laplace transform $\tilde{N}(t)$ for $t\to\infty$
using the Feynman--Kac representation on $N$ (see
Section~\ref{sec:FK}). Their estimate relied on an investigation of
the ``Wiener sausage'' and the machinery of large deviations for
Markov processes developed by these authors. To obtain information
about the behavior of $N(E)$ for $E\searrow 0=E_0$ from the large
$t$ behavior of $\tilde{N}(t)$ one uses Tauberian theorems
\cite{BGT,Bruijn}. This technique was already used by Pastur
\cite{Pastur3,Benderskij} and developed in \cite{Nakao,Fukushima2}
and recently in \cite{Metzger}.

Donsker and Varadhan \cite{DV} needed in their proof of (\ref{ds})
that the single site potential $f$ decays faster than
$(1+\no{x})^{-(d+2)}$. They  asked whether this condition is
necessary for the result (Lifshitz tails) or just necessary for
\emph{their proof}. It was Pastur (\cite{Pastur4}) who
observed that, in fact, the Lifshitz asymptotic is qualitatively
changed if $f$ has long range tails, i.e., if $f(x)\sim
C(1+|x|)^{-\alpha}$ for $\alpha<d+2$. Observe that $\alpha>d$ is
necessary for the mere existence of $V_\omega$. Pastur proved the
behavior
\begin{equation}\label{pastur}
    N(E)\sim C_{1}\ e^{-C_{2}(E-E_{0})^{-\frac{d}{\alpha-d}}}
\end{equation}
as $E\searrow E_{0}$ for $d<\alpha<d+2$. We call this behavior
\emph{Pastur tails}. For a disordered system with constant magnetic
field in dimension $d=2$, Pastur tails (\ref{pastur}) were found for
\emph{all} $\alpha>d=2$ in \cite{BHKL}.

These results and more observations of the last several years indicate that
the asymptotics of the integrated density of states even at the
bottom of the spectrum is more complicated than expected. To be more
precise and following the terminology of \cite{PF}, we can
distinguish two qualitatively different behaviors in the low energy
asymptotics of the integrated density of states. For short range
potentials and ``fat'' single site distributions, the asymptotics of
$N(E)$ is determined by the quantum kinetic energy as predicted by
Lifshitz. Hence it is called quantum asymptotics or \emph{quantum
regime}. On the other hand, for long range potentials or ``thin''
single site distributions, the leading asymptotics of the integrated
density of states is determined by the potential, i.e., by classical
effects. This situation is called the \emph{classical regime}.

We will discuss these phenomena in this section. We start with the
short range case (quantum regime). The proof of Lifshitz tails we
present here is based on spectral theoretic arguments  close to
Lifshitz's original heuristics (see Section \ref{sec:shortrange}).

We then discuss the long range case (classical regime) (Section
\ref{sec:longrange}) to some extent, including recent results
\cite{KiWa1} of single site potentials with anisotropic decay
resulting in a mixed classical-quantum regime (Section~\ref{sec:aniso}).

Classical and quantum behavior of the integrated density of states
and the transition between the two regimes is best understood for
the Anderson model. The approach of \cite{Metzger} combines spectral
theoretic and path integral methods. We will present this in Section~\ref{cominatio}.

Lifshitz predicted the behavior (\ref{os}) and (\ref{ds}) not only
at the bottom of the spectrum but also for any band edge of the
spectrum. To distinguish these two cases, we will speak of {\it
internal Lifshitz tails} in the latter case. Investigating Lifshitz
behavior at internal band edges turns out to be much more
complicated than at the bottom of the spectrum.
In fact, already the investigation of periodic potentials at internal
band edges is extremely complicated. We will discuss
internal Lifshitz tails (following \cite{Klopp3, KW1, KW2}) in
Section~\ref{sec:internal}.

Finally, we will look at random Schr\"odinger operators with
magnetic fields in Section~\ref{sec:magnet}.

\subsection{Lifshitz Tails: Quantum Case\label{sec:shortrange}}\label{DNB0}

\subsubsection{Statement of the main result}
The aim of this subsection is a proof of Lifshitz behavior
close to his original heuristics and without heavy machinery. We
will prove the quantum asymptotics in (\ref{ds}) for short range
single site potentials and ``fat'' single site distributions. We will
make no attempt to reach high
generality but rather emphasize the strategy of the proof.

As before, we consider random alloy-type potentials of the form
\begin{equation}\label{def:alloy}
 V_\omega(x)=\sum_{i\in\Z^d}\;q_i(\omega)\,f(x-i).
 \end{equation}
We assume that the random variables $q_i$ are independent and
identically distributed with a common probability distribution
$P_0$. We suppose that the support of $P_0$ is compact and contains
at least two points.

As always, we also suppose that the single site potential $f$ is
non-negative, bounded and decays at infinity as fast as
$\no{x}^{-(d+\varepsilon)}$. The technique we are going to present
allows us to treat local singularities of $f$. (We refer to \cite{KS,
KiWa1} for details.)

To ensure Lifshitz tails in the sense of (\ref{ds}), we need two conditions:

\smallskip
\noindent\emph{Assumption 1:} Define $q_{min}=\inf\,\supp(P_0)$. We
assume that
\begin{equation}\label{ass:fat}
P_0\big([q_{\min}, q_{\min}+\varepsilon)\big)~\ge~C\,\varepsilon^N
\end{equation}
for some $C$, $N$ and all $\varepsilon>0$ small.

Condition (\ref{ass:fat}) means that the distribution $P_0$ is ``fat''
at the bottom of its support. Note that this condition is, in
particular, satisfied if $P_0$ has an atom at $q_{\min}$, i.e.,  if
$P_0(\{q_{\min}\})>0$.

The second condition we need is precisely the ``short range''
condition already encountered by Donsker and Varadhan \cite{DV}.

\smallskip
\noindent\emph{Assumption 2:}
\begin{equation}\label{ass:sr}
    f(x) \le C\,(1+\no{x})^{-(d+2)}.
\end{equation}

We are ready to formulate the main result of this subsection.
\begin{theorem} \label{satz}
If Assumptions (1) and (2) are satisfied, we have
\begin{equation}\label{result}
   \lim_{E\searrow E_{0}}{\ln (-\ln N(E))\over \ln (E-E_{0})}=-\frac{d}{2}.
\end{equation}
\end{theorem}

Observe that equation (\ref{result}) is a weak form of
Lifshitz's original conjecture (\ref{ds}). In their work \cite{DV},
Donsker and Varadhan proved the stronger form for the Poisson potential
\begin{equation}\label{lifdv}
\lim_{E\searrow 0}{\ln N(E)\over E^{-d/2}}=-C_{d}
\end{equation}
where $C_{d}$ is a (computable) positive constant.

Both the short range condition (Assumption 2) and the fatness
condition (Assumption 1)
 turn out to be necessary for the above result, as we will see later. For example, if
 the single site potential $f$ decays substantially slower than required in the short
 range condition the integrated density of states decays \emph{faster}
 than in the (\ref{result}).

We define the \emph{Lifshitz exponent} $\gamma$ by
\begin{equation}\label{def:lifexp}
    \gamma = \lim_{E\searrow E_{0}}{\ln (-\ln N(E))\over \ln
    (E-E_{0})}
\end{equation}
whenever this limit exists. With this notation we may rephrase
(\ref{result}) as $\gamma=-\,d/2$.
The Lifshitz exponent for periodic potentials is $0$.\\[2mm]

\subsubsection{Strategy of the proof\label{ssec:strategy}}
The proof of Theorem \ref{satz} consists of an upper and a
lower bound. The next subsection will provide us with the tools we
need for these bounds.

It will turn out that the bounds are easier and more natural for
positive random potentials. Therefore, we will split the random
potential in a periodic and a positive random part
\begin{align}
    V_\omega(x) &= \sum_{i\in\Z^d}\;q_{\min}\,f(x-i)+
    \sum_{i\in\Z^d}\;(q_{i}(\omega)-q_{\min})\,f(x-i)\notag\\
    &= V_{\text{per}}(x) + \tilde{V}_\omega(x)\label{split}.
\end{align}
We will subsume the periodic potential under the kinetic energy and
denote the positive random potential $\tilde{V}_\omega$ in a slight
abuse of notation again by $V_\omega$. Thus we have
\begin{equation}\label{split2}
    H_\omega = H_1 + V_\omega
\end{equation}
with $H_1=H_0+V_{\text{per}}$ a Hamiltonian with periodic ``background''
potential $V_{per}$ and
\begin{equation}
    V_\omega = \sum_{i\in\Z^d}\;q_i(\omega)\,f(x-i)
\end{equation}
where the independent $q_i\ge 0$ have a common probability
distribution
$P_0$ with $0=\inf\,(\supp P_0)$.

For the upper bound below, we need information about the two lowest
eigenvalues of $H_1$ restricted to a box. If $V_{\text{per}}\equiv 0$, these
eigenvalues can be computed explicitly. However, if
$V_{\text{per}}\not\equiv 0$, we need a careful analysis of periodic
operators. This was done in \cite{KSJFA} and \cite{Mezincescu2}.
Here we restrict ourselves to the case
$V_{\text{per}}\equiv 0$, avoiding some technical complications. Note that
this implies $E_0=\inf\,(\sigma(H_\omega))\;=\;0$. We refer the
reader to the papers \cite{KS} and \cite{Mezincescu2} for the
general case. We also remark that \cite{KiWa1} contains an
extension of the approach presented here that works for Poisson
potentials and various other potentials as well. \vspace{2mm}

\subsubsection{The Dirichlet--Neumann bracketing}
 The first step in the proof is to bound the integrated density of states from above and from below
using the Dirichlet--Neumann bracketing as in Corollary \ref{DNB}.
We have
\begin{equation}\label{DNB1}
    \frac{1}{\,|\Lambda_L|\,}\;
  \E\,\big(N(H_{\omega\,{\Lambda_{L}}}^{D}, E)\,\big) \le  N(E) \le
\frac{1}{\,|\Lambda_L|\,}\;
\E\,\big(N(H_{\omega\,{\Lambda_{L}}}^{N}, E)\big).
\end{equation}
The side length $L$ of the cube $\Lambda_{L}$ will be chosen later
in an $E$-dependent way when we send $E$ to $E_{0}$. We estimate the
right-hand side of (\ref{DNB1})   by
\begin{align*}
\E (N(H_{\omega\,\Lambda_{L}}^{N},E)
&=\int N(H_{\omega\,\Lambda_{L}}^{N},E)~d\,\P \\
&= \int_{E_{1}(H_{\omega\,{\Lambda_{L}}}^{N})\le
E}N(H_{\omega\,{\Lambda_{L}}}^N,E)~d\,\P
+ \int_{E_{1}(H_{\omega\,{\Lambda_{L}}}^{N})> E}N(H_{\omega\,{\Lambda_{L}}}^N,E)~d\,\P \\
&\le \P \,(E_{1}(H_{\omega\,{\Lambda_{L}}}^{N})\le E) ~N({H_0\,}_{\Lambda_{L}}^{N},E\,).
\end{align*}
With $ N({H_0\,}_{\Lambda_{L}}^{N},E\,) \le
(C_{1}+C_{2}E)^{d/2}|\Lambda|$  following from Weyl asymptotics, we
get for $0\le E\le 1$ the estimate
\begin{equation}\label{DNB2}
\frac{1}{\,|\Lambda_L|\,}\;
\P\,\big(E_{1}(H_{\omega\,{\Lambda_{L}}}^{D})\le E\big) \le
N(E) \le
 C\, \P\,\big(E_{1}(H_{\omega\,{\Lambda_{L}}}^{N})\le E\big).
\end{equation}
The problem now is to find upper and lower bounds such that after
taking the double logarithm, the left- and the right-hand side of
(\ref{DNB2}) coincide asymptotically. In general the upper bounds
are more difficult than the lower bounds. To prove the lower bound
we only have to ``guess'' a good test function, whereas for the upper
bounds, one has to prove that all eigenfunctions for energies in
$[0,E)$
roughly behave the same way.

It is an astonishing fact that the lower bound from (\ref{DNB2}) in
all known cases leads to the \emph{asymptotically correct} behavior
of the integrated density of states, a fact emphasized by Pastur.

\smallskip
\subsubsection{The lower bound}\label{sec:lowb}
For simplicity we   restrict ourselves to  single site
potentials $f$ with $\textnormal{supp}\,
f\subset\Lambda_{\frac{1}{2}}$ so that $f(\cdot-i)$ and $f(\cdot-j)$
do not overlap for $i\not=j$. The necessary changes for the general
case will become clear when we discuss long range potentials $f$.

By the Neumann--Dirichlet bracketing in (\ref{DNB2}), we have for
arbitrary $L$ and any $\psi\in D(\Delta_{\Lambda_L}^D)$ with
$||\psi||_{L^2(\Lambda_L)}=1$,
\begin{align}
\nonumber N(E)&\geq |\Lambda_L|^{-1}\P
\left(E_1({H_\omega}_{\Lambda_L}^D)<E\right) \\
\label{est:lbgen} &\geq |\Lambda_L|^{-1}\P
\left(\langle\psi,H_\omega\psi\rangle_{L^2(\Lambda_L)}<E\right).
\end{align}
A natural   choice of $\psi$ for (\ref{est:lbgen}) seems to be the
ground state $\psi_0$ of $-\Delta_{\Lambda_L}^N$,
$\psi_0(x)\equiv|\Lambda_L|^{-\frac{1}{2}}$. Unfortunately,  this
function does not obey Dirichlet boundary conditions and is therefore
not admissible for (\ref{est:lbgen}).

This problem can be circumvented by multiplying  $\psi_0$ by a
function which is zero at the boundary of $\Lambda_0$. To do so, let
us take $\chi\in C^\infty(\R^d)$, supp$\chi\subset\Lambda_1$,
$\chi(x)=1$ on $\Lambda_{\frac{1}{2}}$ and $0\leq\chi(x)\leq1$. We
set $\chi_L(x)=\chi(\frac{x}{L})$ and
$\psi_L(x)=\chi_L(x)\psi_0(x)$. Then $\psi_L\in
D({\Delta}_{\Lambda_L}^D)$, $||\psi_L||\geq\frac{1}{2}$ and
\begin{align*}
\langle\psi_L,H_\omega\psi_L
\rangle &\leq \langle\psi_0,H_\omega\psi_0 \rangle+CL^{-2}\\
&= |\Lambda_L|^{-1}\int_{\Lambda_L}V_\omega(x)dx+CL^{-2}.
\end{align*}
Note that the ``error term'' $L^{-2}$ is due to the influence of the
kinetic energy (a second-order differential operator). Inserting in
(\ref{est:lbgen}) we get
\begin{align}
\nonumber N(E)&\geq |\Lambda_L|^{-1}~\P
\left(|\Lambda_L|^{-1}\int_{\Lambda_L}V_\omega(x)<E-CL^{-2}\right)\\
\label{est:lbN} &\geq |\Lambda_L|^{-1}\;\P
\biggl(|\Lambda_L|^{-1}\;\|\,f\,\|_1\;\sum_{i\in\Lambda_L}q_i(\omega)<E-CL^{-2}\biggr).
\end{align}
In principle, we can choose $L$ as we like. However, if $E<CL^{-2}$
estimate (\ref{est:lbN}) becomes useless. So it seems reasonable to
choose   $L=\beta E^{-\frac{1}{2}}$ and we obtain
\begin{align*}
(\ref{est:lbN})&\geq |\Lambda_L|^{-1}\;\P \biggl(|\Lambda_L|^{-1}\sum_{i\in\Lambda_L}q_i(\omega)<\tilde{C}E\biggr)\\
&\geq |\Lambda_L|^{-1}\;\P\,\big(q_0<\tilde{C}\,E\big)^{L^d}\\
&\ge C_1 E^{-\frac{d}{2}} \big(C\,E^N\big)^{C_2 E^{-\frac{d}{2}}}.
\end{align*}
Thus we conclude
\begin{equation}
\lim_{E\searrow 0}\,\frac{\,\ln(-\ln(N(E)))\,}{\ln E}\ge
-\frac{d}{2}.
\end{equation}

\smallskip
\subsubsection{The upper bound}
The strategy to prove the upper bounds of $\P\,
(E_1(H_{\omega_{\Lambda_{L}}}^{N})$ $< E)$ in (\ref{DNB2}) can be
divided in two parts. The first step is to find a \emph{lower} bound
for $E_1(H_{\omega_{\Lambda_{L}}}^{D}) $ in such a way that it is
possible to control the influence of the random potential. This is
done by an application of Temple's inequality. The second step is to
balance between the size of $\Lambda_L$ and the probability of a
random potential such that most of the potential values are small.

We start by stating Temple's inequality for the reader's
convenience. A proof can be found, e.g., in \cite{ReedSimon4}.

\begin{theorem}[Temple's inequality] Suppose $H$ is a self-adjoint operator, bounded
below which has discrete spectrum and denote by $E_n(H),n=1,2,\dots$
its eigenvalues $($in increasing order, counting multiplicity$)$. If
$\mu \le E_2(H)$ and $\psi \in { D}\,(H) $ with $\Vert \psi \Vert =
1 $ satisfying $ \langle \psi , H \, \psi \rangle  < \mu $, then
$$
E_1(H) \ge \langle \psi ,H \psi \rangle  - { {\langle \psi ,H^2 \, \psi \rangle  -
{\langle \psi ,H\psi \rangle }^2} \over {\mu - \langle \psi ,H\,
\psi \rangle } }.
$$
\end{theorem}

To apply Temple's inequality, we set
$E_2{(-\Delta_{\Lambda_L}^N):=\mu\le E_2(H_\omega}_{\Lambda_L}^N)$.
Note that by direct computation,
$E_1{(-\Delta_{\Lambda_L}^N)}=0=E_0$  and
$E_2{(-\Delta_{\Lambda_L}^N)} \sim L^{-2}$. These facts require a
careful analysis if there is a periodic background potential as in
(\ref{split}) and (\ref{split2}); see \cite{KSJFA}.

Next we need a good approximation $\psi$ of the ground state of
${H_\omega}_{\Lambda_L}^N $. This is done by  choosing $\psi$ to be
the ground state  of $-\Delta_{\Lambda_L}^N$, which is intuitively
close to the correct ground state for small $E$. The function $\psi$
is given by $\psi(x)= \,\no{\LL}^{-1/2}$.

To apply Temple's inequality we have to ensure that with the above
choice, $\langle \psi , H \, \psi \rangle  < \mu \approx c\,
L^{-2}$. We force this to happen by changing the coupling constants
$q_i$ to $\tilde{q}_i = \min \, (q_i(\omega),\alpha L^{-2})$ with a
suitable $\alpha>0$, small enough. If $\tilde{H}$ denotes the
corresponding operator, we have $E_1(H)\ge E_1(\tilde{H})$. An
application of Temple's inequality to $\tilde{H}$ and an elementary
calculation yield the following lemma.

\begin{lemma} \label{coupl}
\begin{equation}
E_1({H_\omega}^N_{\Lambda_L})   \ge  {1 \over 2} ~ {1\over\vert
\Lambda_L \vert}~ \sum_{i\in \Lambda_L}\tilde q_i(\omega).
\end{equation}
\end{lemma}

A consequence of the lemma above is the intuitively
convincing estimate
\begin{align}\label{coupl2}
  \P\,\big(E_{1}({H_{\omega\,}}_{\Lambda_{L}}^{N})<  E\,\big)
& \le   \P\,\biggl({1\over \vert \Lambda_{L}\vert}
\sum_{i\in \Lambda_{L}}\tilde q_{i}(\omega)\le 2\,E\,\biggr)\\
&\le \P\,\biggl(\;{1\over L^{d}}\;\;\sum_{\vert i\vert_\infty \le\,
{L/2}}\tilde q_{i}(\omega)\le 2\,E\biggr).\label{prob:ld}
\end{align}
The expression (\ref{prob:ld}) for $E$ small very much resembles a
large deviation probability which would lead to a bound
exponentially small in the volume term $L^d$. At first sight,
Cramer's theorem, a  result of the theory of large deviation, seems
to be applicable (see, e.g., \cite{Dembo,Deuschel, H}).

However, there is a complication here: To obtain a large deviation
event in (\ref{prob:ld}) we need that $\E(\tilde{q}_i)<E$. Thus,
if we set $L=L(E)= \beta E^{-1/2}$ with $\beta>0$ small, the event
(\ref{prob:ld}) is, indeed, a large deviation event and we obtain
the following bound.

\begin{lemma}
$$  \P\,\biggl({1\over \vert \Lambda_{L}\vert}
\sum_{i\in \Lambda_{L}}\tilde q_{i}(\omega)\le E\,\biggr) \le
C_{1}~e^{-\,C_{2}\;L^{d}}$$ for $E$ close enough to zero and $L\le
\beta\,E^{-1/2}$.
\end{lemma}
\noindent Combining the results above, (\ref{DNB2}) and (\ref{coupl2}),
we have proven
$$N(E)\le C_{1}^{-C_{2}E^{-d/2}}.$$

\subsubsection{Final remarks}
The idea of using Neumann--Dirichlet bracketing to prove
Lifshitz tails first appeared in \cite{KM4}. It was carried over to
the discrete Anderson model by Simon \cite{Simon2} who streamlined
it at the same time. The proof was  extended to more general
alloy-type potentials by Kirsch and Simon \cite{KS}, who still
needed reflection symmetry of $f$. Mezincescu \cite{Mezincescu2}
modified the upper bound by introducing other boundary conditions to
get rid of this extra assumption. We refer to \cite{KiWa1} for a
rather general proof using these techniques.

\subsection{Long Range Single Site Potentials: A ``Classical'' Case\label{sec:longrange}}
In this section we turn to an example of classical
behavior of the integrated density of states near
$0=\inf(\sigma(H_\omega))$, in the sense of Section~\ref{intro},
namely, to long range single site potential $f$.

The upper bound on $N(E)$ is  easier than for the short range case.
While there is a subtle interplay between the kinetic energy and the
potential in the short range case ($f(x)\leq|x|^{-(d+2)}$), it is
the potential energy alone that determines the leading behavior of
$N(E)$ ($E\searrow0$) in the long
range case.

\smallskip
\noindent \emph{Assumption:} In this section we suppose that
\begin{equation}\label{def:lr}
    \frac{c}{\,(1+\no{x})^\alpha\,} \le f(x) \le \frac{C}{\,(1+\no{x})^\alpha\,}.
\end{equation}

\begin{theorem}\label{th:LTlr} Assume \eqref{ass:fat}. If
\eqref{def:lr} holds for an $\alpha$ with $d<\alpha<d+2$, then
\begin{equation}
\lim_{E\searrow E_0}\frac{\ln(-\ln N(E))}{\ln
E}=-\frac{d}{\alpha-d}.
\end{equation}
\end{theorem}

In the terminology of (\ref{def:lifexp}), Theorem~\ref{th:LTlr} states that the
Lifshitz exponent for the long range case ($\alpha<\,d+2$) is $d/(\alpha-d)$.

\begin{proof}
To simplify the argument, we assume as in the short range case, there
is no periodic background potential and $q_{\min}=0$. Consequently,
$E_0=0$. We start with the upper bound and estimate
\begin{eqnarray*}
E_1({H_\omega}_{\Lambda_1}^N) \geq
\inf_{x\in\Lambda_1}\;\sum_{i\in\Z^d}q_i\frac{c}{(x+|i|)^\alpha}.
\end{eqnarray*}
Hence
\begin{align}
\nonumber N(E)&\leq C_1\;\P\,\big(E_1({H_\omega}_{\Lambda_1}^N)<E\big)\\
\nonumber &\leq C_1\;\P\,\biggl(\, \sum_{i\in\Z^d}q_i\frac{C_2}{(1+|i|)^\alpha}<E\biggr)\\
\nonumber &\leq C_1\;\P\biggl(\, \sum_{|i|\leq
L}q_i\frac{C'}{L^\alpha}<E\biggr)\\
\label{eq:stern} &\leq C_1\;\P\biggl(\frac{1}{L^d}\sum_{|i|\leq
L}q_i<C_3EL^{\alpha-d}\biggr).
\end{align}
We choose $L=\frac{\beta}{C_3}E^{-\frac{1}{\alpha-d}}$ ($\beta$
small). Hence
\[
(\ref{eq:stern})\leq  C_1~\P\biggl(\frac{1}{L^d}\sum_{|i|\leq
L}q_i\leq\beta\biggr).
\]
If $\beta<\frac{1}{2}\,\E(q_0)$, standard large deviation theory
gives
\[
\P\biggl(\frac{1}{L^d}\sum_{|i|\leq
L}q_i\leq\beta\biggr)<e^{-CL^d}=e^{-\tilde{C}E^{-\frac{d}{\alpha-d}}}.
\]

We turn to the lower bound. As in the proof of the lower bound (\ref{est:lbN})
in the previous section,
\begin{align*}
N(E)&\geq \frac{1}{|\Lambda_L|}\P\left(\frac{1}{|\Lambda_L|}\int_{\Lambda_L}
V_\omega (x)dx<E-CL^{-2}\right)\\
&\geq \frac{1}{|\Lambda_L|}\P\biggl(\sum_{i\in\Z^d}q_i
\frac{1}{|\Lambda_L|}\int_{\Lambda_L}f(x-i)dx<E-CL^{-2}\biggr).
\end{align*}
Due to the long range tails of $f$, we cannot ignore the summands
with $|i|$ large. Instead, we estimate
\begin{align*}
\sum_{i\in\Z^d}\,q_i\; &\frac{1}{|\Lambda_L|} \int_{\Lambda_L}f(x-i)\,dx\\
&\leq \frac{1}{|\Lambda_L|}\sum_{|i|_\infty\leq\,2L} q_i\int
f(y)\,dy+\sum_{|i|_\infty>\,2L}q_i\frac{1}{|\Lambda_L|}\int_{\Lambda_L}f(x-i)\,dx\\
&\leq \frac{C_3}{|\Lambda_{2L}|}\sum_{|i|_\infty\leq\,2L}q_i+q_{\max}\sum_{|i|_\infty>\,2L}\frac{1}{|\Lambda_L|}\int_{\Lambda_L
}f(x-i)\,dx,
\end{align*}
where $q_{\max}=\sup(\textnormal{supp}P_0)$, $P_0$ being the
distribution of the $q_i$. We estimate
\begin{align*}
\sum_{|i|>2L}\frac{1}{|\Lambda_L|}\int_{\Lambda_L}f(x-i)dx
& \leq C_4\sum_{|i|>2L}\frac{1}{|\Lambda_L|}\int_{\Lambda_L}\frac{1}{|x-i|^\alpha}dx \\
&\leq C_5\sum_{|i|>L} \frac{1}{|i|^\alpha}\\
&\leq C_6\frac{1}{L^{\alpha-d}}.
\end{align*}
Thus we obtain
\begin{eqnarray}\label{est:lbkp}
N(E)\geq\frac{1}{|\Lambda_L|}\P\biggl(\frac{1}{|\Lambda_{2L}|}\sum_{|i|\leq
2L}q_i\leq C_7E-C_8L^{-2}-C_9L^{_(\alpha -d)}  \biggr).
\end{eqnarray}
As the derivation shows, the $L^{-2}$ term comes from manipulating
the kinetic energy, while the $L^{-(\alpha-d)}$ term is due to the
potential energy. Note that for $\alpha>d+2$, the term $L^{-2}$
(kinetic energy contribution) dominates in (\ref{est:lbkp}). In this
case we can therefore redo the estimates of Section~\ref{sec:lowb}
and obtain a lower bound as we got there.
However, for $\alpha<d+2$, the term $L^{-(\alpha-d)}$ wins out in
(\ref{est:lbkp}). Remember, this term is due to the potential energy
distribution. We obtain
\[
N(E)\geq\frac{1}{|\Lambda_L|}\P\biggl(\frac{1}{|\Lambda_{2L}|}\sum_{|i|\leq
2L}q_i\leq C_7E-C_{10}L^{-(\alpha-d)}\biggr).
\]
This time, $E$ has to be bigger than $L^{-(\alpha-d)}$; more
precisely, $E\geq C_{11}L^{-(\alpha-d)}$. Hence $L=C_{12}E^{-\frac{1}{\alpha-d}}$ so
\begin{align*}
N(E)&\geq \frac{1}{|\Lambda_L|}\P\left(q_i=0\textnormal{ for
}|i|\leq2L \right)\\
&\geq \frac{1}{|\Lambda_L|} e^{-C_{13}L^d}\\
&\geq  C_{14}E^{\frac{d}{\alpha-d}}\
e^{-C_{15}E^{-\frac{d}{\alpha-d}}}.
\qedhere
\end{align*}
\end{proof}

\subsection{Anisotropic Single Site Potentials\label{sec:aniso}}
Recently, Theorems \ref{satz} and \ref{th:LTlr} were
generalized to single site potentials $f$ decaying in an anisotropic
way at infinity (\cite{KiWa1}). Let us write
$x\in\R^d=\R^{d_1}\times\R^{d_2}$ as $x=(x_1,x_2)$,
$x_1\in\R^{d_1}$, $x_2\in\R^{d_2}$ and suppose that
\begin{eqnarray}\label{pot:decai}
\frac{a}{|x_1|^{\alpha_1}+|x_2|^{\alpha_2}}\leq f(x)\leq
\frac{b}{|x_1|^{\alpha_1}+|x_2|^{\alpha_2}}
\end{eqnarray}
for $|x_1|,|x_2|\geq1$, and define $V_\omega$ in the usual way
\begin{eqnarray}\label{pot:at}
V_\omega(x)=\sum_{i\in\Z^d}q_i f(x-i).
\end{eqnarray}
Let us define $\gamma_i=\frac{d_i}{\alpha_i}$ and
$\gamma=\gamma_1+\gamma_2$. Then the sum in (\ref{pot:at}) converges
(absolutely) if $\gamma<1$.

In \cite{KiWa1} the authors prove that there is Lifshitz behavior
of $N(E)$ for potential as in (\ref{pot:decai}) and (\ref{pot:at})
in the sense that the Lifshitz exponent $\eta$, defined by
\[
\eta=\lim_{E\searrow E_0}\frac{\ln|\ln(N(E))|}{\ln E}
\]
exists ($E_0=\inf\sigma(H_\omega)$). $\eta$ depends on the
exponents $\alpha_i$, of course.
If both
\begin{equation}\label{case:qq}
\frac{\gamma_1}{1-\gamma}\leq\frac{d_1}{2}\quad\textnormal{and}\quad\frac{\gamma_2}{1-\gamma}\leq\frac{d_2}{2}
\end{equation}
we obtain the ``quantum'' exponent:
\[
\eta=-\frac{d}{2}.
\]

Observe that (\ref{case:qq}) reduce to the condition
$\alpha\le d+2$ for the isotropic case $\alpha_1=\alpha_2=\alpha$.
If
\begin{eqnarray}\label{case:cc}
\frac{\gamma_1}{1-\gamma}>\frac{d_1}{2}\quad\textnormal{and}\quad\frac{\gamma_2}{1-\gamma}>\frac{d_2}{2}
\end{eqnarray}
we are in the ``classical'' case both in the $d_1$- and $d_2$-directions. Then
\[
\eta=-\frac{\gamma}{1-\gamma}.
\]
The third case
\begin{equation}\label{case:qc}
\frac{\gamma_1}{1-\gamma}\leq\frac{d_1}{2}\quad\textnormal{and}\quad\frac{\gamma_2}{1-\gamma}>\frac{d_2}{2}
\end{equation}
is new compared to the isotropic case. It is, in a sense, a mixed
quantum-classical case. The Lifshitz exponent is given by
\[
\eta=-\frac{d_1}{2}-\frac{\gamma_2}{1-\gamma}.
\]
We note that the $d_1$-direction and the $d_2$-direction ``influence
each other'' in a rather sophisticated way. In \cite{KiWa1} these
results are proved for alloy-type potentials as well as for Poisson
(and related) models. We summarize:

\begin{theorem}[\cite{KiWa1}]\label{th:aniso}
Suppose \eqref{ass:fat} and \eqref{pot:decai} hold. Set
$\gamma_i= d_i/\alpha_i$ and $\gamma=\gamma_1+\gamma_2$. Then
the Lifshitz exponent $\eta$ is given by
\begin{equation}
\eta = -\max\, \biggl\{\frac{d_1}{2},\frac{\gamma_1}{1-\gamma}\biggr\} -
\max\, \biggl\{\frac{d_2}{2},\frac{\gamma_2}{1-\gamma}\biggr\} .
\end{equation}
\end{theorem}

\subsection{Path Integral Methods and the  Transition Between Quantum  and
Classical Regime with Respect to the Single Site Measure\label{cominatio}}
We start this section with two observations which indicate
that the asymptotic behavior of the integrated density of states
depends qualitatively on the distribution $P_0$ of the $q_i$.

Let us first assume that the ``fatness'' condition (\ref{ass:fat}) is
satisfied in the strongest form, namely, $\inf\,\supp(P_0)=0$ and
$P(q_{0}=0)=a > 0$.
An inspection of the proofs in Section~\ref{DNB0} shows that for
this case we have actually proven
\begin{align}\label{ab1}
\liminf\;{\ln N(E)\over (E-E_{0})^{-d/2}} &\ge -C_{1} , \\
\label{ab2}
\limsup\;{\ln N(E)\over (E-E_{0})^{-d/2}}&\le -C_{2}
\end{align}
with   $C_{1},C_{2}> 0$.  If instead $P(q_{0}=0)=0$  (but still
$P(q_{0} > \epsilon )\ge B\epsilon^{n}$) the lower bound requires a
logarithmic correction
\begin{equation}\label{ab3}
\liminf\;{\ln N(E)\over (E-E_{0})^{-d/2}\vert\ln
(E-E_{0})\vert}\ge -C_{1}.
\end{equation}
The second observation concerns unbounded single site measures
$P_0$. In \cite{Klopp5} it  is proved that both the classical and
the quantum regime can occur for the discrete, unbounded Anderson
model and more general matrix operators.  Depending on the single site
measure, collective phenomena may occur similar to those we encountered above.
In other situations,  the single site measure alone determines the behavior
of the integrated density of states.

It seems difficult to understand the mechanisms causing the
transition from quantum  to classical regime with respect to the
single site measure by using the spectral analytic approach close to
Lifshitz's original intuition.

The first approach to prove Lifshitz tails is based on the Donsker--Varadhan
technique (see also Section \ref{intro}). This method to
compute the Laplace transform of the integrated density of states in
the limit $t\rightarrow \infty$ is a far-reaching generalization of
the Laplace method known from classical analysis. The starting point
is the path integral representation
\begin{displaymath}\label{fkn2}
 \tilde N(t)=\E\times
\E_{0,0}^{t,0}\bigl[e^{-\int_{0}^{t}V_{\omega}(b(s))ds}\bigr] .
\end{displaymath}
The Donsker--Varadhan technique is based on a  large deviation
principle satisfied by the product probability measure $d\P\times
d\P_{0,0}^{t,0}$ combining the random potential and   the Brownian
motion. In an informal sense, it makes it possible to quantize   the
asymptotic probability of a Brownian particle to stay most  of its
lifetime in a pocket with a favorable configuration of potential
values. Using the large deviation principle, one can  balance between
favorable configurations and their small probability by applying
Varadhan's lemma. Last but not least, given the large time
asymptotics  of the Laplace transform $\tilde N(t)$, one can
reconstruct the Lifshitz tail behavior using Tauber theory.

The Donsker--Varadhan technique was worked out by Nakao \cite{Nakao}
for the Poisson model with $f\ge 0$. He proved
$$
\lim_{E\searrow 0}{\ln N(E)\over E^{-d/2}}=-C_{d}
$$
where $C_{d}$ is a (computable) positive constant.  In the 1990's
 Lifshitz asymptotics became a starting point for
stochastic analysis of diffusion in random media. We mention the
work of Sznitman (see, e.g, \cite{Sznitman1,Sznitman2,Sznitman3,Sznitman4})
in the continuous case, especially for Poisson potentials, and in the
discrete context the moment analysis for the so-called parabolic
Anderson model (PAM) (see, e.g., \cite{bk1,GM1,GM2,GM3}).
Here Brownian motion has to be replaced by the
continuous time Markov chain  generated by the discrete Laplacian.

The phenomenology described at the beginning of this subsection was
also observed in the moment analysis of the parabolic Anderson model
starting in \cite{GM1}. As we will see,  the case of the double
exponential distribution discussed in \cite{GM2} can be interpreted
as the borderline between the quantum and the classical regime. The
paper \cite{bk1} clarified the discrepancy between the lower and the
upper bounds in (\ref{ab1}), (\ref{ab2}) and (\ref{ab3}). Still, a
general principle explaining the transition from quantum  to
classical regime with respect to the single site measure was not
formulated.

We want to systemize the phenomenology discussed above in the
following theorem taken from \cite{Metzger}. To combine the bounded
and the unbounded case,  we assume that the cumulant generating
function is finite, e.g.,
\begin{equation}\label{cumul1}
    G(t):=\log\,\E\,\big(\exp(-tV_\omega(0))\big)<\infty
\end{equation}
 for all $t\ge0$. Furthermore, we apply the Legendre transformation to define the rate function
\begin{equation}\label{cumul2}
    I(E):= \sup_{t>0}[Et-G(t)]
\end{equation}
and for $t,\lambda>0$,   we set
\begin{equation}\label{cumul3}
    S(\lambda,t):=(\lambda t)^{-1}G(\lambda t)- t^{-1}G( t).
\end{equation}
Informally the scale function $S(\lambda,t)$   measures   the change
of the cumulant generating function after rescaling the time.

\begin{theorem}\label{satz1}
We consider the discrete Anderson tight binding operator $h_\omega
=h_0+V_\omega$ and set $E_0=\inf \sigma(h_\omega)$. Suppose $G(t)
<\infty $ for all $t>0$. Then we distinguish the following four
cases:

\smallskip
\noindent{\rm{(i)}}  Let $ S(\lambda,t) \;\sim\;  c\:( \lambda^{\rho}-1)
t^{\rho}$ with $c,\rho >0$. Then the IDS behaves in the limit
$E\rightarrow E_0=-\infty$ like
\begin{align*}
 \log(N(E))=- I(E+2d+o(1))(1+o(1)).
\end{align*}

\smallskip
\noindent{\rm{(ii)}} In the case $S(\lambda,t) \sim  c\,\log(\lambda)$
with $\rho=0$ and $c >0$, we have in the limit $E\rightarrow E_0=-\infty$,
\begin{align*}
 -KI(E+C_2)(1+o(1))\le\log(N(E))\le -I(E+C_1)(1+o(1))
\end{align*}
with $K>0$,
\begin{align*}
C_1 =  - 2    \sin^2\left(\frac{ \pi}{2}\;\frac{1}{c^{-1/2}
+1}\right) + \frac{ 1 }{4}\:c\log(c)
\end{align*}
and
\begin{align*}
C_2 =   K(d )\min\biggl[  -  c +    c\log(c^{-1}), \max\biggl[
1-4\exp(-Kc) ,\frac{a(d) }{4}\biggr]\biggr]   .
\end{align*}

\smallskip
\noindent{\rm{(iii)}} In the case $S(\lambda,t) \sim   c\,(1-
\lambda^{\rho} ) t^{\rho}$ with $-1< \rho <0$ and $c >0$, the
IDS behaves in the limit $E\searrow
E_0=0$ like
\begin{align*}
 -K_1 E^{-1/2(d-2\rho^{-1}(\rho+1))}(1+o(1))
 &\le\log(N(E))\\
 &\le -K_2  E^{-1/2(d-2\rho^{-1}(\rho+1))}(1+o(1)).
\end{align*}

\smallskip
\noindent{\rm{(iv)}} In the case $S(\lambda,t) \sim   -c\,(\lambda
t)^{-1} \log(t)$ with $c >0$, we have in the limit $E\searrow
E_0=0$,
\begin{align*}
 -K_1 E^{-d/2 }\log(E)(1+o(1))
 &\le\log(N(E))\\
 &\le -K_2  E^{-d/2  }\log(E)(1+o(1)).
\end{align*}
\end{theorem}

The scaling assumption $ S(\lambda,t) \sim  c\,(
\lambda^{\rho}-1) t^{\rho}$ with $c,\rho >0$ in the first case
corresponds to ``fat'' unbounded single site distributions; the
behavior of the integrated density of states is classical. The
second case represents the double exponential case, while in the
third situation the single site distribution is bounded, but very
thin. The fourth case corresponds to relatively fat single site
distributions studied in \cite{KM1}.

Although, by now, there are results covering   a lot  of  possible
single site distributions,  there seems to be no systematic approach
known to explain this phenomenology. Furthermore there exists two
relatively different approaches as discussed above. A first step to
combine the functional analytic and the path integral approach as
well as to systemize the known results with respect to the single
site distribution seem to be   \cite{Metzger,Metzger1}. In contrast
to the   direct analysis of the operator $H_\omega$ in Section~\ref{DNB0},
but in analogy to the path integral methods, one is
interested in the large time behavior of the semigroup
$\exp(-h_\omega t)$.   The Lifshitz asymptotics follows by an
application of (modified) Tauber theorems  \cite{Metzger2}.

The first step of the argument in \cite{Metzger} is to restrict
$h_\omega$ to a (time-dependent) box $\Lambda=\Lambda_t(0)$ by
introducing discrete Dirichlet boundary conditions and to approximate
$\E[\exp(-h_\omega t)(0,0)]$ in the limit $t\rightarrow \infty$ by
\begin{equation}\label{approx}
\E[\exp(-E_1(h_\omega^{\Lambda}) t) ]  = \;  \E \biggl[\: \sup_{p\in
M_1(\Lambda) }\exp
\left(-t\;\left[\left(\sqrt{p}|h_0^{\Lambda}\sqrt{p}
\right)+\left(\sqrt{p}|V(\omega)\sqrt{p}
\right)\right]\right)\biggr].
\end{equation}
Here $E_1(h_\omega^{\Lambda}) =\inf \sigma(h_\omega^{\Lambda})$ is
the principal eigenvalue of $h_\omega^{\Lambda}$ and $M_1(\Lambda)$
is the set of   probability measures   on $\Lambda$. Equation
(\ref{approx}) is a consequence of the min-max principle and the
nonnegativity of the ground state. It is the starting point to find
upper and lower bounds of the Laplace
transform of the IDS.

To illustrate the central effect explaining the transition from the quantum
mechanical to the classical regime, we want to sketch the very elementary proof of the
lower bounds starting from (\ref{approx}). The first step is to interchange
the expectation value and the supremum
\begin{align}
\nonumber     \E \left[    \exp(-E_1 (h_\omega^{\Lambda}) t)  \right]
       &=  \E \biggl[\: \sup_{p\in  M_1(\Lambda) }\exp
\left(-t\;\left[\left(\sqrt{p}|h_0^{\Lambda}\sqrt{p}
\right)+\left(\sqrt{p}|V(\omega)\sqrt{p} \right)\right]\right)\biggr] \\
\nonumber       &\ge  \sup_{p\in  M_1(\Lambda) }  \exp \left(-t\;
\left(\sqrt{p}|h_0^{\Lambda}\sqrt{p}\right)\right)\E \biggl[
\exp \biggl(-\sum_{x \in \Lambda} p(x)V_\omega(x)t\biggr)\biggr]\\
\label{est1}                &= \sup_{p\in  M_1(\Lambda) }
               \exp\biggl(-t\; \left(\sqrt{p}|h_0^{\Lambda}\sqrt{p}\right)+\sum_{x \in \Lambda}
               G(p(x)t)\biggr).
\end{align}
The second step is to define a subset $D\subset M_1(\Lambda)$ of
relatively uniform probability distributions concentrated on a
subvolume of $\Lambda$. With the side length L  of $\Lambda$, we set
$1\le l \le L$ and $\Lambda_l:= \{x\in \Z^d: |x|_\infty \le l\}$.
The ground state of the discrete Laplacian $h_0^{\Lambda_l}$
restricted to $\Lambda_l$ with Dirichlet boundary conditions is
given by
\begin{align}
  \phi_l &\colon  \Lambda_l \rightarrow [0,\infty), \\
  \phi_l &= \prod_{j=1}^d \left(\frac{2}{l+1}\right)^{1/2}
  \sin\left(\frac{x_j\pi}{l+1}\right)
\end{align}
and the  corresponding principal eigenvalue of $h_0^{\Lambda_l}$ is
\begin{equation}\label{prineig}
    E_1(h_0^{\Lambda_l})= 2d\biggl(1-\cos\biggl(\frac{\pi}{l+1}\biggr)\biggr).
\end{equation}
The subset $D\subset M_1(\Lambda)$ is then defined by
\begin{equation}\label{defd}
    D:= \{\phi_l^2:1\le l\le L\}.
\end{equation}
Restricting the estimate (\ref{est1}) to $D$, we get
\begin{align*}
\E\,\big(\exp\left(-t\, \right. & \left.E_1\left(h_\omega^{\Lambda} \right)\right)\big) \\
& \ge  \; \exp\biggl( G(t) + t \; \sup_{p\in  M_1(\Lambda) }
\biggl( \;-(\sqrt{p}|h_0^{\Lambda}\sqrt{p})\;+\;
  \sum_{x\in \Lambda}p(x) S(p(x),t)\biggr)\biggr)\\
& \ge  \; \exp\biggl( G(t) + t \; \sup_{p\in    D } \biggl(
\;-(\sqrt{p}|h_0^{\Lambda}\sqrt{p})\;+\;
  \sum_{x\in \Lambda}p(x) S(p(x),t)\biggr)\biggr).
\end{align*}
Using the definition of $S(\lambda,t)$, the convexity of the
cumulant generating function $G(t)$ and the Jensen inequality, we can
estimate for $p\in  M_1(\Lambda_l)\subset  M_1(\Lambda)$,
\begin{align*}
  \sum_{x\in \Lambda}p(x) S(p(x),t)  \; =& \;  \sum_{x\in
\Lambda_l}p(x)\biggl(\frac{G(p(x)t)}{p(x)t}-\frac{G(t)}{t}\biggr) \\
=& \; t^{-1}l^d\biggl(l^{-d}\sum_{x\in \Lambda_l}G(p(x) t)\biggr) - t^{-1}G(t)\\
\ge& \; t^{-1}l^d G \biggl(l^{-d}t\sum_{x\in \Lambda_l} p(x)   \biggr) - t^{-1}G(t)\\
=& \; \frac{G( l^{-d}t)}{l^{-d}t}-\frac{G(t)}{t}\\
=& \;     S(l^{-d},t).
\end{align*}
So the uniform distribution on   $\Lambda_l $ minimizes $\sum_{x\in
\Lambda}p(x) S(p(x),t)$ with respect to the variation over  $p\in
M_1(\Lambda_l)$. We have
\[
\E\,\big(\exp\left(-t\; E_1\left(H_{\Lambda}^D (\omega)\right)\right)\big)
\ge \exp\biggl( G(t) + t \sup_{1\le l\le L} \biggl( - 4d
\sin^2\biggl(\frac{ \pi}{2}\;\frac{1}{l+1}\biggr) +
S(l^{-d},t)\biggr)\biggr)
\]
and the only remaining problem is to maximize
\begin{align*}
     - 4dt  \sin^2 &\biggl(\frac{ \pi}{2}\;\frac{1}{l+1}\biggr)  +
tS(l^{-d},t) \\
& \sim  - 4dt   \sin^2\biggl(\frac{\pi}{2}\;\frac{1}{l+1}\biggr) +
\begin{cases}
 c\:( l^{-d\rho}-1) t^{\rho+1}  &  \rho >0\\
  -cdt \:\log(l)          & \rho =0\\
 c\:(1- l^{-d\rho} )  t^{\rho+1}  & -1<\rho <0   \\
 -c\: l^d   \log(t)
\end{cases}
\end{align*}
with respect to $l$. The exponent of the time $t$ in the scaling
expression is responsible for the occurrence of the classical or the
quantum regime.  In the case   $\rho >0$, the scaling term increases
faster in $t$ than the linear time dependence in the diffusion term.
Consequently, the maximum  will be asymptotically $l=1$. This
corresponds to the classical regime. In the case $-1<\rho <0$  as
well as in the fourth situation, the scaling term is sublinear and
the diffusion term is dominating. Like in Section~\ref{DNB0},  a
collective behavior of potential values is necessary  and we are in
the quantum regime.   In the case $\rho =0$, the diffusion and the
scaling term are both linear in time. So the optimal peak size
depends strongly on the constants. This is the borderline between
the classical and the quantum   regime. It corresponds
to the double exponential distribution.

The upper bounds are  much more complicated. It is not possible to
interchange the supremum and the expectation with respect to the
random potential. Moreover, we have to to estimate (\ref{approx})
for all $p\in M_1(\Lambda)$. The first problem is solvable by a
variant of the ordinary Laplace method. The second problem is
attacked using the convexity of the cumulant generating function
$G(t)$ and ideas from spectral geometry. For details, we refer to
\cite{Metzger} and \cite{Metzger1}.

\subsection{Internal Band Edges\label{sec:internal}}
Lifshitz predicted the ``Lifshitz behavior'' not only
for the bottom of the spectrum but also for other band edges. We
refer to this phenomenon as ``internal Lifshitz tails.''
Internal Lifshitz tails have been proven for the Anderson model by
Mezincescu \cite{Mezincescu1} and Simon \cite{Simon3}. Their proofs
apparently cannot be translated to the continuum case. In fact, the
band edges of the Anderson model which they can handle are those
coming from gaps in supp$\,P_{0}$ together with the boundedness of
the kinetic energy. (To be more precise: Since  for the Anderson
model,  $\Vert\, h_0\Vert \le 4d$ and $h_0\ge 0$, there are gaps in
the spectrum whenever there are gaps in $\supp\,P_{0}$ of length
exceeding $4d$).

One can also handle the case of a point interaction potential in one
dimension, a problem which essentially reduces to a lattice problem.
Formally this potential is given by
$$V_{\omega}=\sum q_{i}\delta (x-i)$$
where $\delta$ is the Dirac-``function.'' This potential is
also known as the random \emph{Kronig--Penney model}. It turns out in
this case that the lower edges do and the upper edges (for $q_{i}\ge
0$) do not show Lifshitz behavior but polynomial behavior of $N$ as
for periodic potentials \cite{KNit}. This is due to the fact that
the upper edges are ``stable boundaries'' in the sense of \cite{PF}.
The case of general one-dimensional alloy-type potentials was treated in
\cite{Mezincescu3}.

The multidimensional case is by far more difficult. The reason is
mainly that \emph{periodic} potentials are much less well
understood in higher dimensions. For example, it is not true in
general that bands are parabolic, as is the case for $d=1$ and for
the ground state band in arbitrary dimension.

The paper \cite{Klopp3} marks a breakthrough in this topic. Klopp
uses the method of approximation by periodic potentials. Compared to
Dirichlet--Neumann bracketing, one loses monotonicity, a property
which was very useful above. However, Klopp manages to prove an
exponential convergence rate for the periodic approximations.

As mentioned above, not so much is known about the behavior of the
band functions (of the periodic operators)  at internal band edges.
In fact, Klopp has to make assumptions on the behavior of the
integrated density of states for the \emph{periodic} operator.

Like Lifshitz tails at the bottom of the spectrum internal Lifshitz
tails can be used as an
input for a localization proof \cite{Veselic1}.

We consider an alloy-type potential with a continuous single site
potential  $f\ge 0$, not identically equal to $0$, with decay
\begin{equation}
f(x)\le C\,(1+\no{x})^{-(d+2+\varepsilon)}.
\end{equation}
The random coupling constants are independent and have a common
probability distribution $P_0$ with $q_{\min}=\inf\supp P_0$.  We set
$V_{\text{per}}=\sum_{i\in\Z^d}\,q_{\min}\,f(x-i)$ and denote the integrated
density of states of $H_{\text{per}}=H_0+V_{\text{per}}$ by $N_{\text{per}}(E)$.
Furthermore, we suppose that $E_-$ is a lower band edge of $H_{\text{per}}$,
i.e., $E_-\in\sigma(H_{\text{per}})$, but
$(E_--a,E_-)\cap\sigma(H_{\text{per}})=\emptyset$.

It is reasonable to assume that generically $N_{\text{per}}$ behaves like
$(E-E_-)^{d/2}$, for $E\searrow E_-$, as it would for a unique
parabolic band. In fact, this behavior is known in one dimension and
for the bottom of the spectrum in arbitrary dimension. However, it
is not clear that this is true in general, even not generically
(see, however, \cite{KR}).

Thus, we have to \emph{assume} such a behavior of $N_{\text{per}}$:

 \smallskip
\noindent\emph{Assumption:} Suppose $E_-$ is a lower band edge of
$H_{\text{per}}$. We assume that
\begin{equation}\label{ass:quadr}
\lim_{E\searrow
E_-}\,\frac{\,\ln\,\big(\,N_{\text{per}}(E)-N_{\text{per}}(E_-)\,\big)\,}{\ln\,
(\,E-E_-\,)}=-\frac{d}{2}\, .
\end{equation}
Under this assumption, the main result of \cite{Klopp3} is:

\begin{theorem}[Klopp] If assumption \eqref{ass:quadr} holds, then
\begin{equation}\label{th:Lifinternal}
\lim_{E\searrow
E_-}\,\frac{\ln\,\Big(-\ln\,\big(\,N(E)-N(E_-)\,\big)\,\Big)}{\ln\,
(\,E-E_-\,)}=-\frac{d}{2}\, .
\end{equation}
\end{theorem}

For the case $d=2$, one has more information about the
periodic operators \cite{KW2}. In particular, there is always
exponential decay of the integrated  density of states at
band edges. We refer to the review \cite{Klopp7} for an introduction
and further results.

\subsection{Lifshitz Tails for Surface Potentials}
In this section we consider surface potentials of the form
\begin{equation}\label{dec:surf}
V^s_\omega(x_1,x_2) =
\sum_{i_1\in\Z^{d_1}}q_{i_1}(\omega)f(x_1-i_1,x_2)
\end{equation}
and suppose we have some spectrum below $0$. This is the case if
$q_{\min}=\inf\,\supp\,P_0$ is negative enough. Note that for
$d_2\le2$, there is negative spectrum as soon as $q_{\min}$ is
negative. For $d_2\ge3$, there is a threshold $\gamma>0$ such that
the spectrum starts at $0$ if $q_{\min}\ge -\gamma$ and there is
negative spectrum if $q_{\min}< -\gamma$.

We are going to investigate Lifshitz tails for surface potentials
with $E_0<0$. Below the bulk spectrum (which starts at $0$), the
density of surface states is positive, hence a measure. We may
therefore define the integrated density of surface states $N_s(E)$
to be the corresponding distribution function.

As before, we decompose the potential into a non-random background
potential and a positive random potential
\begin{align}
V^s_\omega(x_1,x_2) &= \sum_{i_1\in\Z^{d_1}}q_{i_1}(\omega)f(x_1-i_1,x_2)\notag\\
&= \sum_{i_1\in\Z^{d_1}}q_{\min}\,f(x_1-i_1,x_2)
+ \sum_{i_1\in\Z^{d_1}}(q_{i_1}-q_{\min})\,f(x_1-i_1,x_2)\notag\\
&= V^s_{\textnormal{sp}\,}(x)~+~\tilde{V}_\omega^s(x).
\end{align}
The Neumann--Dirichlet bracketing technique goes through for this
case as soon as we have sufficient knowledge about the background
operator $H_1=H_0+V_{\textnormal{sp}}^s$. Since we want $E_0<0$
---which makes the bottom of the spectrum ``surface spectrum''---there is
no case $V_{\textnormal{sp}}^s=0$ here. Moreover, the background
potential $V_{\textnormal{sp}}^s$ is only periodic for the
$d_1$-directions, but decays perpendicular to them.

The analysis of those partially periodic potentials and the Lifshitz
estimates for surface potentials were done in \cite{KiWa2} for the
continuous case. We assume that
$$
P_0\big( [q_{\min}, q_{\min}+\varepsilon) \,\big)\,\ge C\,\varepsilon^N
$$
and
$$
0\;\le\; f(x_1, x_2)\;\le\;f_0\;(1+\no{x_1})^{-(d_1+2)}.
$$
We also assume that $f(x_1,x_2)$ decays uniformly in
$x_2$-directions. Then we have:

\begin{theorem}
If $E_0<0$, then
\begin{equation}
    \lim_{E\searrow E_0}\;\frac{\,\ln\big(-\ln(N_s(E))\big)\,}{\ln\,
    (E-E_0)}=-\frac{d_1}{2}\, .
\end{equation}
\end{theorem}

There is also an analogous theorem for long range $f$. The
paper \cite{KK} proves Lifshitz tails for surface potentials in the
discrete setting by fairly different techniques. This paper also contains
an analysis at the energy $E=0$, i.e., for surface corrections
to the bulk Lifshitz tails.

\subsection{Lifshitz Tails for Random Landau Hamiltonians\label{sec:magnet}}
We turn to the density of states for operators of the form
$$
H_\omega=H_0(B)+V_\omega
$$
with a constant magnetic field $B>0$ and a non-negative random
potential $V_\omega$. (For a careful definition of the density of
states and some basics, see \cite{HLMW1,HLMW2,Warzel}.)

We discuss the \emph{two-dimensional} case first. The Landau
Hamiltonian $H_0(B)$ is given by
$$
H_0(B)=\biggl(i\frac{\partial}{\partial
x_1}-\frac{1}{2}Bx_2\biggr)^2+\biggl(i\frac{\partial}{\partial
x_2}+\frac{1}{2}Bx_1\biggr)^2.
$$
$H_0(B)$ has a pure point spectrum for $B\not=0$ and $d=2$. In
fact, the eigenvalues are given by the ``Landau levels''  $(2n+1)B;~~
n\in\N$ and all Landau levels are infinitely degenerate.  One
possible choice of the ground state is
\begin{equation}\label{maggrstate}
\psi_0(x)=\frac{B}{\pi}\;e^{-\frac{B}{2}|x|^2}
\end{equation}
which will play a major role below. For $V_\omega$ we take a Poisson
potential or an alloy-type potential with $q_{\min}=0$. In this case,
the bottom $E_0$ of the spectrum of $H_\omega$ is given by the
lowest Landau level, which is $B$. In \cite{BHKL} the authors proved
for the Poisson model the following result.

\begin{theorem}\label{th:lifmag1}
If $B\not=0$ and
\begin{equation}
\frac{C_1}{(1+|x|)^\alpha}\leq f(x)\leq\frac{C_2}{(1+|x|)^\alpha},
\end{equation}
the Lifshitz exponent $\eta$ for $H_0(B)+V_\omega$ is given by
\begin{equation}
\eta=\frac{2}{2-\alpha}~~~ \biggl(=\frac{d}{d-\alpha}\biggr)
\end{equation}
for \emph{all} $\alpha>d$.
\end{theorem}

This means that, according to our classification above, we
are always in the classical case for $d=2$ and constant magnetic
field.

\begin{proof}
We sketch the lower bound only and restrict ourselves to the
alloy-type case. As usual we have to estimate
$$
\mathbb{P}(E_1({H_\omega}_{\Lambda_L}^D) < E_0+E)
$$
from below. This time we have a ground state $\psi_0$ for $H_0(B)$ which
is $L^2$. We modify $\psi_0$ near the boundary of
$\Lambda_L$ to make it satisfy Dirichlet boundary conditions. Due to
the (super-)exponential decay of $\psi_0$, the error we make is of
the order (at most) $e^{-C_0\,L^2}$. Thus
\begin{align*}
\mathbb{P}(E_1({H_\omega}_{\Lambda_L}^D) < E_0+E)
&\geq \mathbb{P}\biggl(\int V_\omega(x)|\psi_0|^2 dx < E-e^{-C_0\,L^2}\biggr)\\
&\geq  \mathbb{P}\biggl(\int V_\omega(x) dx < C_1E-e^{-C_2\,L^2}\biggr) \\
&=  (*) \ .
\end{align*}
At this point we can literally repeat the estimates in the proof of
Theorem \ref{th:LTlr} and obtain in analogy to (\ref{est:lbkp})
\begin{equation}\label{est:magn}
(*) \geq  \mathbb{P}(\frac{1}{|\Lambda_{2L}|} \sum_{|i|\leq 2L} q_i
\leq C_1 E - C_2 e^{-cL^2}-C_3 L^{-(\alpha - d)} ) \ .
\end{equation}
The only difference to the previous case is the error term due to
the kinetic energy. In Theorem \ref{th:LTlr} it was of the order
$L^{-2}$ causing the different behavior for $\alpha\geq d+2$ and
$\alpha < d$. In (\ref{est:magn}) the error term is exponentially
small, thus being negligible with respect to the potential term
$L^{-(\alpha -d)}$ for \emph{all} $\alpha$. Consequently, we may
choose $E\sim L^{\frac{1}{\alpha - d}}$. By a large deviation
estimate, we obtain
\begin{align*}
(\ref{est:magn}) &\geq  \mathbb{P}(q_i \leq C'E)^{|\Lambda_{2L}|} \\
 &\geq  M e^{-\tilde{C}L^d} \\
 &=  M e^{-\tilde{\tilde{C}} E^{-\frac{d}{\alpha - d}}}\, .
\qedhere
\end{align*}
\end{proof}

Theorem \ref{th:lifmag1} implies that for compactly
supported $f$,
the integrated density of states $N(E)$ decays subexponentially.

Erd\"os \cite{Erdoes1, Erdoes2} proved that it decays, in fact,
polynomially. Erd\"os' proof is based on a careful estimate of the
Laplace transform of $N$. It uses an analog of the Feynman--Kac
formula for magnetic Schr\"odinger operators, the Feynman--Kac--Ito
formula (see \cite{Simon5}). There are a couple of complications in
the Feynman--Kac expression of $\tilde{N}$ due to the magnetic field.
The most serious one is the fact that the integrand is no longer
positive but rather oscillating.

Erd\"os' proof is done for the Poisson model. Recently, Klopp and
Raikov \cite{KRai} found a completely different approach based on
the approximation by periodic potentials. Their proof works for
alloy-type potentials. Moreover, the latter paper contains results
for internal Lifshitz tails as well as for the case of an additional
periodic background potential.

We state Erd\"os' result:

\begin{theorem}\label{th:lifmag2}
If $H_\omega=H_0(B)+V_\omega$ with a Poisson random potential
$V_\omega$ and if the single site potential $f\ge 0$ has compact
support, then
\begin{equation}\label{th:erdoes}
    \lim_{E\searrow E_0}\;\frac{\,\ln N(E)\,}{\ln
    E}~=~-\,\frac{\pi}{B}.
\end{equation}
\end{theorem}

The reasoning we gave in the proof of Theorem~\ref{th:lifmag1}
can be used to prove a lower bound for Theorem~\ref{th:lifmag2} as well.
In fact, the bound suggests that the
borderline between the two kinds of behavior is given by Gaussian
single site potentials. This was actually proved in \cite{HLW} and
\cite{Erdoes2}.

We have seen in this section that random Landau Hamiltonians may
have unusual Lifshitz behavior compared to the nonmagnetic case.
This originates in the fact that the lowest Landau band is flat,
collapsing into one point. It is well known that this never happens
for ($B=0$ and) periodic \emph{scalar} potential (see
\cite{ReedSimon4}).

It is reasonable to expect that the flatness of the ground state
Landau band is removed by an additional periodic scalar potential,
at least generically. If this is true, we would certainly expect
``normal'' Lifshitz behavior for such operators, including those with
alloy-type potentials and $q_{\min}\not=0$. There are results in this
direction in \cite{KRai}.

\smallskip
We turn to the \emph{three-dimensional} case with homogeneous
magnetic field. We consider the Hamiltonian
\begin{eqnarray*}
H_0(B)=\biggl(i\frac{\partial}{\partial
x_1}-\frac{1}{2}Bx_2\biggr)^2+\biggl(i\frac{\partial}{\partial
x_2}+\frac{1}{2}Bx_1\biggr)^2-\frac{\partial^{\,2}}{\partial\,
{x_3}^{2}}\, .
\end{eqnarray*}
In $d=3$, the magnetic field itself introduces an anisotropy. The two
space dimensions perpendicular to the magnetic field $(x_1, x_2)$
will be denoted by $x_\perp$, the direction $x_3$ parallel to the
field by $x_\parallel$. If we add an anisotropic Anderson (or
Poisson) potential with a single site potential $f$ obeying
\begin{eqnarray}\label{pot:dec}
\frac{a}{|x_\perp|^{\alpha_1}+|x_\parallel|^{\alpha_2}} \leq
f(x) \leq  \frac{b}{|x_\perp|^{\alpha_1}+|x_\parallel|^{\alpha_2}}\, ,
\end{eqnarray}
we will have Lifshitz behavior that resembles the results of
Section~\ref{sec:aniso}, except that in the $\perp$-direction the behavior
is always \emph{classical}.

Following the conventions of Section~\ref{sec:aniso}, we define
\begin{equation}
\gamma_1=\frac{2}{\alpha_1}\;,\;\;  \gamma_2=\frac{1}{\alpha_2}
\;\textnormal{ and }\;\gamma=\gamma_1+\gamma_2.
\end{equation}
If $f$ has compact support in $x_\perp$-direction, we set
$\gamma_1=0$. To have the potential well defined, we need $\gamma<1$.

\begin{theorem} The Lifshitz exponent $\eta$ for $H_0(B)+V_\omega$ in dimension
$d=3$ is given by
\begin{equation}
\eta = \frac{\gamma_1}{1-\gamma} + \max\,\biggl\{\frac{1}{2}\;,\;
\frac{\gamma_2}{1-\gamma}\biggr\}.
\end{equation}
\end{theorem}
\noindent This theorem was proven for $f$ with compact support in
\cite{Warzel}. The case of $\gamma_2>\frac{1}{2}$ is considered in
\cite{HKW} and the mixed classical-quantum case is taken from
\cite{KLW}.

We remark  that the lower bound given above (in connection with
Theorem~\ref{th:lifmag1}) can be used in this case as well. As a
test function in this bound, we use
\begin{equation}
\Psi_0(x_\perp,x_\parallel) = \psi_0(x_\perp)\,\phi_0(x_\parallel)
\end{equation}
where $\psi_0$ is the (2-d) ground state (\ref{maggrstate}) and
$\phi_0(x_\parallel)= L^{-1/2} $ is the $\parallel$-ground state. As
before $\Psi_0$ has to be cut down to zero near the boundary of the
cube $\LL$. This gives an error term of the order $e^{-L^2}$ in the
$\perp$-direction and an error term of the order
$L^{-2}$ in the $\parallel$-direction.

For further references about magnetic Lifshitz tails, we refer to
\cite{Warzel} and to \cite{LW}.

There are also results on the integrated density of states and on
Lifshitz tails for \emph{random} magnetic fields. We refer to the
works of Ueki and Nakamura \cite{Ueki, Ueki2, Ueki3, Nakamura2,
Nakamura3} and the references given there.

\subsection{Percolation Models\label{sec:percolation}} \noindent
Recently the integrated density of states for Laplacians on
percolation graphs has been investigated. We consider bond
percolation on $\Z^d$ with $d\ge2$. This means that we remove bonds
in the graph $\Z^d$ independently with probability $1-p$ for
$0<p<1$. The resulting random graph is called the \emph{percolation
graph} and is denoted by $\mathcal{G}_p$ or
simply by $\mathcal{G}$.

For $p$ small ($p<p_c$), all connected components (``clusters'') of
$\mathcal{G}_p$ are finite almost surely. For $p>p_c$ there is a
unique infinite cluster (see, e.g., \cite{Grimmett}).

We consider the Laplacian on this graph which is a random operator,
due to the randomness of the underlying graph. For a general
discussion of percolation Hamiltonians, we refer to
\cite{Veselic2}.

Actually, there are various Laplacians on $\mathcal{G}$ due to
different boundary conditions. To define these operators, we start
with the adjacency operator $A_{\mathcal{G}}$ defined by the matrix
elements with   $A_{\mathcal{G}}(i,j)=1$ if $|i-j|=1$, $i,j\in
\mathcal{G}$ and $A_{\mathcal{G}}=0$ otherwise. For any
$i\in\mathcal{G}$, we let $d_{\mathcal{G}}(i)$ denote the number of
sites in $\mathcal{G}$ to which $i$ is connected. We denote by
$D_{\mathcal{G}}$ the diagonal matrix with entries
$d_{\mathcal{G}}(i)$ on the diagonal.  The Neumann Laplacian
$L_{\mathcal{G}}^N$ is defined by
\begin{equation}
L_{\mathcal{G}}^N = D_{\mathcal{G}} - A_{\mathcal{G}}.
\end{equation}
The Neumann Laplacian is the intrinsic Laplacian of the graph
$\mathcal{G}$. In a sense it ``ignores'' the embedding of
$\mathcal{G}$ into $\Z^d$. The Dirichlet Laplacian
$L_{\mathcal{G}}^D$ is defined by
\begin{equation}
L_{\mathcal{G}}^D = D_{\mathcal{G}}  +
2\,(2d-D_{\mathcal{G}}) - A_{\mathcal{G}}.
\end{equation}
On its diagonal this operator counts the connections to neighboring
sites \emph{in} $\mathcal{G}$ once and the (lost) connections to
sites in $\Z^d\setminus\mathcal{G}$ twice. It was an observation of
Simon \cite{Simon2} that the above operator is a good analog of the
Dirichlet Laplacian for subgraphs of $\Z^d$. See also
\cite{KMueller} or \cite{Kirsch5} for a discussion of
boundary conditions for discrete Laplacians.

Both for $L_{\mathcal{G}}^N$ and for $L_{\mathcal{G}}^D$, a
integrated density of states can be defined in analogy to Theorem
\ref{T1}. We call them $N_{\mathcal{G}}^N(E)$ and
$N_{\mathcal{G}}^D(E)$, respectively.

Since there are infinitely many clusters containing just one point,
the Neumann Laplacian $L_{\mathcal{G}}^N$ has an eigenvalue of
infinite multiplicity at the bottom $E=0$ of its spectrum. This
causes the integrated density of states to jump at that energy.
Hence
$$\nu_{\mathcal{G}}^N(\{0\}) = N_{\mathcal{G}}^N(0)-N_{\mathcal{G}}^N(0-)
= N_{\mathcal{G}}^N(0)~\not= 0.$$
\begin{theorem}
{\rm{(1)}} If $p<p_c$, we have
\begin{equation}
   \lim_{E\searrow 0}\,\frac{\ln\, \Big(-\ln \big(N_{\mathcal{G}}^N(E)-N_{\mathcal{G}}^N(0)\big)\Big)}{\ln
   (E)}=-\frac{1}{2}.
\end{equation}

\smallskip
\noindent{\rm{(2)}}  If $p>p_c$, we have
\begin{equation}
\lim_{E\searrow 0}\,\frac{\ln
\big(N_{\mathcal{G}}^N(E)-N_{\mathcal{G}}^N(0)\big)}{\ln
(E)}=\frac{d}{2}.
\end{equation}

\smallskip
\noindent{\rm{(3)}} For arbitrary $0<p<1$, we have
\begin{equation}
\lim_{E\searrow 0}\,\frac{\ln\, \Big(-\ln
N_{\mathcal{G}}^D(E)\Big)}{\ln (E)}=-\frac{d}{2}.
\end{equation}
\end{theorem}

Part (1) of this theorem, as well as part (3) for $p<p_c$,
was proven in \cite{KMueller}; part (2) and part (3) are  taken from
\cite{MS}.\\

The perhaps surprising behavior of $N_{\mathcal{G}}^N(E)$ for
$p<p_c$ is due to the fact that long one-dimensional chains
dominate. They lead to small eigenvalues for the Neumann Laplacian
since they ``don't know'' they are actually in $\Z^d$. For the
Dirichlet Laplacians these one-dimensional chains have rather high
eigenvalues due to the additional diagonal term. In fact, they
dominate the scene at $E=4d$, the top of the spectrum. This can be
seen by a symmetry argument.

Part (2) of the theorem comes from the fact that for $p>p_c$ the
leading behavior of $N_{\mathcal{G}}^N$ comes from the infinite
cluster. In a sense, the infinite cluster looks rather
$d$-dimensional. The proof in \cite{MS} of this part relies on a
celebrated paper by Barlow \cite{Barlow}.

\section{Regularity of the Integrated Density of States\label{sec:regular}}
\subsection{Introduction}
\noindent In this final section we discuss regularity properties of
the integrated density of states. So far we have seen that the
density of states measure $\nu$ is a positive Borel measure with a
distribution function $N(E)$.

Of course, the name integrated \emph{density} of states suggests
that $\nu$ (resp.\ $N$) should have a density $n(E)$ in the sense
\begin{align}
N(E)&= \int_{E'\,\le\,E}\;n(E')\;dE',\\
\nu([a,b])&= \int_{[a,b]}\;n(E')\;dE'.
\end{align}
We will, indeed, prove this for the \emph{Anderson model} under
certain assumptions on $P_0$, the probability distribution of the
random potential. The continuous case is more complicated. For this
case we will state only some key results and refer the reader to the
literature. A good account is the review article by Veselic
\cite{Veselic}. However, we remark that there are important
developments after this survey; we mention especially the recent preprint \cite{CHKnew}.

We will see that the integrated density of states does \emph{not}
always have a density. We will therefore also look at weaker
regularity properties of $N$. We are especially interested in the
question whether the function $N$ is continuous, which is the same
as $\nu(\{E\})=0$ for all $E$. It turns out that this is always the
case for the Anderson model. However, for the continuous case, no
such result is known.

\subsection{Continuity of the Integrated Density of States}
The results of this section are valid for general ergodic
operators on $\ell^2(\Z^d)$. Analogous results for $\R^d$
have not been proven so far.
The most general result in this context is the following theorem
proved by Craig and Simon \cite{CS1} in a somewhat stronger form
(see Theorem~\ref{th:logcont}).

\begin{theorem}[\cite{CS1,DS2}]\label{cont}
Let $\{V_\omega(n)\colon n\in \Z^d\}$ be an ergodic stationary real valued
random potential satisfying
\begin{equation}
    \E[\log(1+|V_\omega(0)|] <\infty.
\end{equation}
Then the integrated density of states of the Anderson operator
$h_\omega=h_0+V_\omega$  is a continuous function, i.e.,
$\nu(\{\lambda\})=0$ for all $\lambda\in\R$.
\end{theorem}

 This result can be proven with an elementary argument by
Delyon and Souillard  \cite{DS2,CFKS}   using a kind of ``unique
continuation'' property of discrete Schr\"odinger operators. We use
this idea in the proof of the following lemma to show that no
eigenspace can be sufficiently degenerated to produce a jump of
the integrated density of states.

Let us denote by $\mu_H(\cdot)$ the projection-valued spectral
measure of $H$, i.e., $\mu_H(A)=\chi_A(H)$; in particular,
$\mu_H(\{\lambda\}\,)$ is the projector onto the eigenspace of $H$
with respect to $\lambda$.

\begin{lemma}\label{L1}
We have
 \begin{equation*}
   \dim \big(\,\chi_{\Lambda_L}\textnormal{Ran}\, \mu_{h_\omega}({\{\lambda\}})\,\big)\leq C\, L^{d-1}.
 \end{equation*}
\end{lemma}

\begin{proof} The set
\begin{equation*}
   \Lambda^{(2)}_L=\biggl\{i\in\Lambda_L|\quad \max_{\nu=1, \ldots, d}\;|i_\nu|=L~
\textnormal{ or } \max_{\nu=1, \ldots, d}\;|i_\nu|=L-1\biggr\}
\end{equation*}
consists of the two outermost layers of $\Lambda_L$. A solution $u$
of $h_\omega u=\lambda u$ is uniquely determined inside $\Lambda_L$
by its values on $\Lambda^{(2)}_L$. So, the dimension of
$\chi_{\Lambda_L}\big(\textnormal{Ran} \mu_{h_\omega}(\{\lambda\})
\big)$ is at most the number of points in $\Lambda^{(2)}_L$.
\end{proof}

With the lemma, we can prove the theorem.

\begin{proof}[Proof of the Theorem]  By Proposition~\ref{P1} and Theorem~\ref{T1}, we have
 \begin{equation}\label{eq:*}
 \nu(\{\lambda\})=\lim_{L\rightarrow\infty}\frac{1}{(2L+1)^d}
 \tr\,\big(\chi_{\Lambda_L}\,\mu_{h_\omega}(\{\lambda\})\big).
 \end{equation}
If $f_i$ is an orthonormal basis of $\chi_{\Lambda_L}\big(\textnormal{Ran}
\,\mu_{h_\omega}(\{\lambda\})\big)$ and $g_i$ an orthonormal basis of
 $\big(\chi_{\Lambda_L}\big(\textnormal{Ran} \,\mu_{h_\omega}(\{\lambda\})\big)\big)^\bot$ we have,
 noting that $\chi_{\lambda_L}(\text{Ran} \,\mu_{h_\omega}(\{\lambda\}))$ is finite-dimensional,
\begin{align}
     \tr\,\big(\chi_{\Lambda_L}\,\mu_{h_\omega}(\{\lambda\})\big)
   &= \sum_{i\in I}\langle f_i,\chi_{\Lambda_L}\,\mu_{h_\omega}(\{\lambda\})f_i \rangle \nonumber\\
   &\leq \dim\left(\,\chi_{\Lambda_L}\big(\textnormal{Ran}\, \mu_{h_\omega}(\{\lambda\})\big)\right)\\
    &\leq C\, L^{d-1}
\end{align}
hence (\ref{eq:*}) converges to zero and $\nu(\{\lambda\})=0$.
\end{proof}

In \cite{CS1} Craig and Simon prove a stronger result than
just continuity:

\begin{theorem}[Craig--Simon]\label{th:logcont}
Under the assumptions of Theorem~\ref{cont},  the integrated density
of states is locally log-H\"older continuous, e.g., for any positive r
there exists a finite constant $C_r$ such that
\begin{equation}
|N(\lambda)- N(\lambda^{\prime})| \le C_r\;
\frac{1}{\,\no{\log|\lambda -
 \lambda^{\prime}|\,}}
\end{equation}
for $|\lambda|\le r$ and $|\lambda - \lambda^{\prime}|\le 1$.
\end{theorem}

The theorem of Craig and Simon is based on the Thouless formula for a
strip in $\Z^d$. For the Thouless formula in the one-dimensional setting,
we refer to the next subsection.

\subsection{Regularity of the DOS in Dimension One\label{sec:reg1d}}\noindent
There are very powerful techniques to study the Schr\"odinger equation
and its discrete analog in dimension \emph{one} which are
unfortunately restricted to dimension one exclusively. Such
techniques have been successfully applied to random operators as
well to investigate the integrated density of states.

At the heart of most of these techniques lies the reformulation of
the eigenvalue equation of a second-order equation into an initial
value problem for a system of first-order equations. Specifically,
let us look at the eigenvalue equation
\begin{equation}\label{eieq}
   h_\omega = -u(n+1)-u(n-1)+(V_\omega(n)-E)u(n)=0.
  \end{equation}
We define
\begin{equation*}
    U(n)= \biggl(\begin{matrix}u(n+1)\\u(n)\end{matrix}\biggr)
  \end{equation*}
and
\begin{equation*}
    A_n(E)=
    \left(\begin{matrix}V_\omega(n)-E&-1\\1&0\end{matrix}\right).
  \end{equation*}
The function $u(n)$ is  a solution of (\ref{eieq}) if and only if
\begin{equation*}
     U(n+1)= A_{n+1}(E) U(n)
\end{equation*}
for all $n\in \Z$. With the transfer matrix
\begin{equation*}
    \Phi_n(E):= \prod_{i=1}^n A_i(E)
\end{equation*}
the solution of (\ref{eieq}) to the right initial condition
\begin{equation*}
     U(0)= \left(\begin{matrix}u(1)\\u(0)\end{matrix}\right)
\end{equation*}
can be expressed by
\begin{equation*}
    U(n)= \Phi_n(E)U(0).
\end{equation*}
Similarly, it is possible to define the solution to the left. The
spectral theory of the operator $h_\omega$ is encoded in the
matrices $A_n(E)$ (or $\Phi_n(E)$). Note that these matrices belong
to the group $SL(2,\R)$. Thus it should not come as a surprise that
harmonic analysis on $SL(2,\R)$, in explicit or
implicit form, plays a major role in the analysis of $h_\omega$.

The asymptotic behavior of the eigensolutions of
  $h=h_0+V$ is reflected by the Lyapunov exponent  given by
\begin{equation}\label{Lyapunov}
    \gamma(E):= \lim_{N\rightarrow \pm \infty} \ln \|\Phi_n(E)\|.
\end{equation}
This definition of the Lyapunov exponent is well defined by
F\"urstenberg's theorem (\cite{FuKe}, see also \cite{CFKS}). The link
between the Lyapunov exponent and
  the integrated density of states is expressed in the Thouless formula
\begin{equation} \label{Thouless}
    \gamma(E) =\int \log |E-E^{\prime}|\; dN(E^{\prime}).
\end{equation}
This formula from physics (\cite{HerJo,Thou}) was made
rigorous in \cite{AS2}. A simplified proof can be found in
\cite{CS2} or \cite{CFKS}.

The Thouless formula and the obvious fact that $\gamma(E)$ is
non-negative imply that the integrated density of states $N$ is
log-H\"older continuous (in one dimension). This was observed by Craig
and Simon in \cite{CS2}. The proof by the same authors of the
multidimensional analog uses a version of the Thouless formula
for strips in higher dimensions \cite{CS1}.

As mentioned in the previous section, these results hold for general
ergodic potentials on $\Z^d$, they are (at least) close to
optimal in this generality (see also our discussion below).

However, if we assume that the $V_\omega(n)$ are independent (and
identically distributed) much more can be said about the regularity
of the integrated density of states. In the next section, we discuss
Lifshitz continuity in the multidimensional case under assumptions
on the distribution $P_0$ of the random variable $V_\omega(0)$. In
the one-dimensional case we discuss here, we have a fairly complete
picture about the regularity of $N$. For simplicity we assume that
$\supp\,P_0$ is compact.

\begin{definition} We call a function $f$ on $\R$ H\"older continuous of order $\alpha$ if
\begin{equation*}
    |f(x)-f(y)| \le C\;|x-y|^\alpha .
\end{equation*}
\end{definition}

\begin{theorem}\label{th:LePa}
Let $h=h_0+V$ be a discrete random Schr\"odinger operator in dimension
one. Assume $V$ is a sequence of independent, identically distributed
random variables, such that their distribution $P_0$ has compact
support. Then the integrated density of states $N$ of $h$ is H\"older
continuous of some order $\alpha>0$.
\end{theorem}

This theorem is due to \cite{LePa}. The simple case
of the discrete Laplacian $h_0$ (i.e., $V=0$) shows that one cannot
expect more than H\"older continuity in general. For $h_0$ the
integrated density of states $N$ is merely H\"older continuous of
order $\frac12$ at the band edges. The derivative of $N$, which is called
the \emph{density of states}, diverges at $E=\pm 2$. This behavior is
known as van Hove singularities. The Lifshitz behavior of the
integrated density of states seems to suggest that $N$ is
smoother at the band edges for a truly random $V$\!.

One can expect that more regularity for $P_0$ implies more
regularity for $N$. In fact, Simon and Taylor \cite{SimTay} proved
that already a little regularity of $P_0$ implies that $N$ is
$C^\infty$. To state their result, we need the following definition
of a Sobolev space:

\begin{definition}
We say that a function $f\in L^p$ belongs to \emph{$L^p_\alpha$} if
the Fourier transform $\hat{f}$ satisfies: There is a function $g\in
L^p$ such that $\hat{g}(k)=(1+|k|^2)^{\alpha/2}\hat{f}(k)$ is an
$L^p$-function.
\end{definition}

It is not hard to see that the characteristic function
$\chi_{I}$ of a finite interval $I$ is in $L^1_\alpha$ for
$\alpha< \frac12$ (see \cite{SimTay}).

\begin{theorem}\label{th:SiTa}
Let $h=h_0+V$ be a discrete random Schr\"odinger operator in dimension
one. Assume $V$ is a sequence of independent, identically distributed
random variables. If the distribution $P_0$ of $V_\omega(0)$ has a
density $g$ with compact support and such that $g\in L^1_\alpha$ for
some $\alpha>0$, then $N\in C^\infty$.
\end{theorem}

If we do not assume any regularity of $P_0$, Theorem~\ref{th:LePa}
is the best result we can hope for. To see this, let us
look at a Bernoulli distribution for $P_0$. We set
\begin{equation}\label{Bernoulli}
P_0 = p\,\delta_a + (1-p)\delta_b .
\end{equation}
Halperin \cite{Halperin} proved, with some points of rigor clarified
by Simon and Taylor \cite{SimTay}, that:

\begin{theorem}\label{th:Halp}
Assume that the $V_\omega(n)$ are independent with identical
distributions $P_0$ as in \eqref{Bernoulli} with $0<p\le \frac12$. Then
the integrated density of states is \emph{not} H\"older continuous of
any order $\alpha$ larger than
 \begin{equation*}
    \alpha_0=\frac{2\,|\log(1-p)|}{ \textnormal{arccosh} (1+\frac{1}{2}|b-a|)}.
\end{equation*}
\end{theorem}

If $N$ is Lifshitz continuous (H\"older continuous of order
$\alpha=1$), then $N$ has a bounded density $n$, i.e.,
$N(E)=\int_{-\infty}^E\;n(\lambda)\;d\,\lambda$ and vice versa. If
$N$ is H\"older continuous of (strict) order $\alpha<1$, then $N$
may still have a density which then has to be unbounded. So, Theorem~\ref{th:Halp}
does not rule out that the density of states measure $\nu$ is absolutely continuous.
However, it is true that $\nu$ has a
singular continuous component if $|b-a|$ is large.

\begin{theorem}
If $P_0$ is Bernoulli $($as in \eqref{Bernoulli}$)$ and $0<p<1$, then the
density of states measure $\nu$ has a singular continuous component
if $|b-a|$ is large.
\end{theorem}

This theorem was proved in \cite{CKM} following ideas from
\cite{SimTay}. The paper \cite{CKM} contains the first proof of
Anderson localization for the one-dimensional Bernoulli model. The
proof of a singular continuous component of $N$ is based on this
knowledge. The article \cite{MarMi} continues these investigations
and proves that for $|b-a|$ large, the density of
states measure is even purely singular continuous.

For further results concerning the regularity of the DOS of random
Schr\"odinger operators and its discrete analog in dimension  one
see
\cite{Campanino,DSS,Klein3,Kla,Klein2,Klein5,Klein1,Klein5a,Klein4,March,shubin1}.

As the operators with i.i.d.\ potentials, \emph{almost periodic}
operators belong to the class of ergodic operators. In a sense these
two types of operators form the extreme cases within the class of
ergodic operators. There are recent results on regularity of the
integrated density of states for one-dimensional almost periodic
operators by Goldstein and Schlag \cite{gs1,gs2}. We refer to the
review \cite{gs} for details.

\subsection{The Wegner Estimates: Discrete Case}
One motivation in physics to study the integrated density
of states was the hope to use it as an indicator for different
spectral types. The aim was to distinguish pure point spectrum and
continuous spectrum at the mobility edge. In some sense the
conjectures discussed were inconsistent. Some expected  a divergent
density of states at the mobility edge, while
others assumed a vanishing density of states.

In 1981 Wegner \cite{Wegner} put an end to this discussion  by
proving  upper and lower bounds of the density of states in the
discrete setting of the Anderson model. These estimates imply the
density of states neither vanishes nor explodes at a mobility edge
(or anywhere
else).

Wegner's result (more precisely, his upper bound) soon became a
corner piece in the proofs of Anderson localization by the
multiscale analysis method and it still is. To formulate Wegner's
result, we introduce boundary conditions on the lattice. The simplest
boundary condition on the lattice is defining $h^\Lambda_\omega$
through its matrix elements:
\begin{equation}\label{hlambda}
    h^\Lambda_\omega(i,j)=h_\omega(i,j)
\end{equation}
whenever both $i$ and $j$ belong to $\Lambda$. For a discussion of
boundary condition on $\ell^2$, see Simon \cite{Simon2} or the review
\cite{Kirsch5}.

We define
\begin{equation}
    N_{\Lambda}(E)=\#\{\,n\,|\,E_n(h^{\Lambda}_\omega)\le E\} = \tr\big(P_{(-\infty,
    E]}(h^{\Lambda}_\omega)\big).
\end{equation}

\begin{theorem}[Wegner-estimate]{\label{th:Wegner}}  Suppose the measure $P_0$ has
a bounded density $g$, (i.e., $P_0(A)=\int_A g(\lambda)d\lambda, \
||g||_\infty < \infty)$, then
\begin{equation}
\E (N_\Lambda(E+\varepsilon)-N_\Lambda(E-\varepsilon))\leq
2\,||g||_\infty\; |\Lambda|\,\varepsilon.
\end{equation}
\end{theorem}

\begin{remark}
The assumption that $P_0$ has a density cannot be dropped. We have
seen at the end of Section~\ref{sec:reg1d} that the integrated
density of states has a singular continuous component if $P_0$ is a
$($certain$)$ Bernoulli distribution.
\end{remark}
\noindent Before we discuss this estimate, we note two important
consequences. The first concerns the regularity of the density of
states.

\begin{corollary}{\label{co:1}}
Under the assumption of Theorem~\ref{th:Wegner}, the integrated
density of states is absolutely continuous with a bounded density
$n(E)$.
\end{corollary}

\begin{proof}
\begin{align}
N(E+\varepsilon)- N(E-\varepsilon)
&=\lim_{|\Lambda|\rightarrow\infty} \frac{1}{|\Lambda|}
\E (N_\Lambda(E+\varepsilon)-N_\Lambda(E-\varepsilon))\nonumber\\
&\leq C \varepsilon \qquad \textnormal{by Theorem \ref{th:Wegner}.}
\qedhere
\end{align}
\end{proof}

Thus $N(E)=\int_{-\infty}^E n(\lambda)\,d\lambda$. We call
$n(\lambda)$ the \emph{density of states}. Sometimes,  $N$ also is
called the density of states which, we admit, is an abuse of
language.  The second consequence of Theorem~\ref{th:Wegner} is a
key ingredient in proving Anderson localization.

\begin{corollary} {\label{co:2}}
Under the assumptions of Theorem~\ref{th:Wegner}, we have for any $E$
and $\Lambda$,
\begin{equation}
\P (\textnormal{dist}(E,\sigma(h^\Lambda_\omega))<\varepsilon)\leq C
\varepsilon |\Lambda|.
\end{equation}
\end{corollary}

\begin{proof}
  By the Chebyshev inequality, we get
\begin{align}
\P(\textnormal{dist}(E, \sigma(h^\Lambda_\omega))<\varepsilon)
&= \P (N_\Lambda (E+\varepsilon)-N_\Lambda(E-\varepsilon)>1)\nonumber \\
&\leq \E(N_\Lambda(E+\varepsilon)-N_\Lambda(E-\varepsilon)) \nonumber\\
&\leq C \varepsilon |\Lambda| \qquad \textnormal{by Theorem~\ref{th:Wegner}.}
\qedhere
\end{align}
\end{proof}

The first step in the proof of Theorem~\ref{th:Wegner} is
to average the eigenvalues inside the interval $(E-\epsilon,
E+\epsilon)$ with respect to the random potential. To do this we
consider the eigenvalues $E_n(h^\Lambda_\omega)$ as functions of the
arguments $V_i=V_\omega(i)$ with $i\in \Lambda$, i.e.,
$E_n(h^\Lambda_\omega)=E_n(V_i, i\in \Lambda)$. The resulting
estimate is summarized in the following lemma.

\begin{lemma}
With $\varepsilon>0$ let $\varrho$ be a non-decreasing
$C^\infty$-function with $\varrho(\lambda)=1$ for $\lambda >
\varepsilon$, $\varrho(\lambda)=0$ for $\lambda<-\varepsilon$ and
$0\leq \varrho(\lambda)\leq 1$. Then
\[
N(h^\Lambda_\omega, E+\varepsilon)-N(h^\Lambda_\omega,
E-\varepsilon) \le
\sum_{j\in\Lambda}\int_{E-2\varepsilon}^{E+2\varepsilon}
\frac{\partial}{\partial V_j}\,\,
\textnormal{tr}\varrho(h^\Lambda_\omega(\{V_i\})-\eta)d\eta .
\]
\end{lemma}

Now we are in a position to interchange the expectation
of the random potential and the energy integral. Since the random
variables $V_\omega(i)$ are independent and have the common
distribution $d P_0(V_i)$, the expectation $\E$ is just integration
with respect to the product of these distributions. Hence
\begin{align*}
\E(N_\Lambda &(E+\varepsilon) -N_\Lambda(E-\varepsilon))\\
&\leq \E (\sum_{j\in\Lambda}\int_{E-2\varepsilon}^{E+2\varepsilon}
\frac{\partial}{\partial V_j} \textnormal{tr}\
 (\varrho(h^\Lambda_\omega(\{V_i\})-\eta)d\eta)) \\
&=\int_{E-2\varepsilon}^{E+2\varepsilon} d\eta \sum_{j\in\Lambda}
\prod_{i\in\Lambda }\int_\R g(V_i)d V_i \frac{\partial}{\partial
V_j}\, \textnormal{tr}
\varrho(h^\Lambda_\omega(\{V_i\})-\eta)\\
 &\leq ||g||_\infty \int_{E-2\varepsilon}^{E+2\varepsilon} d\eta\;
\sum_{j\in\Lambda}~ \prod_{i\in\Lambda,\ i\neq j}~\int_\R g(V_i)d
V_i \quad\int_{a}^b d V_j \frac{\partial}{\partial V_j}
\textnormal{tr} \varrho(h^\Lambda_\omega(\{V_i\})-\eta) \\
&\le \int_{E-2\varepsilon}^{E+2\varepsilon} d\eta \sum_{j\in\Lambda}
\prod_{i\in\Lambda,\ i\neq j}\int g(V_i)\;d V_i \\
& \qquad \qquad \left\{ \textnormal{tr}\
\varrho(h^\Lambda_\omega(\{V_i\}^{V_j=b})-\eta)-\textnormal{tr}\
\varrho(h^\Lambda_\omega(\{V_i\}^{V_j=a})-\eta)\right\}.
\end{align*}
Above, $a,b$ are such that $\supp g \subset [a,b]$. The
notation $\{V_i\}^{V_j=b}$ means the family $\{V_i\}_{i\in\Lambda}$
with $\tilde{V_i}=V_i$ for $i \neq j$ and $\tilde{V_j}=b$. The
problem is now to estimate   the trace difference. In the discrete
context of the Anderson model, the variation of a potential value at
one site is a rank one perturbation.

\begin{lemma}\label{lem:finrank}
Let $A$ be a self-adjoint operator bounded below with purely discrete
spectrum $E_0 \leq E_1 \leq \dots$ $($where the eigenvalues are repeated
according to multiplicity$)$. If $B$ is a symmetric positive rank one
operator, then $\tilde{A}=A+B$ has eigenvalue $\tilde{E_n}$ with
$E_n\leq\tilde{E_n}\leq{E_{n+1}}$.
\end{lemma}

 Given the lemma, we now continue the proof of the theorem. We
set $A=H_\Lambda(\{V_i\}^{V_j =a})$ and
$\tilde{A}=H_\Lambda(\{V_i\}^{V_j =b})$. Obviously, their difference
is a (positive) rank one operator
\begin{align}
\textnormal{tr}\ \varrho(\tilde{A}-\eta)-\textnormal{tr}\ \varrho(A-\eta)
&= \sum (\varrho(\tilde{E_n}-\eta)-\varrho(E_n-\eta))\nonumber \\
& \leq \sum (\varrho(E_{n+1}-\eta)-\varrho(E_n-\eta))\nonumber \\
& \leq  \sup_{\lambda,\mu}\ \varrho(\lambda)-\varrho(\mu) \nonumber \\
&= 1.
\end{align}

We conclude the proof of Wegner's estimate by proving
Lemma \ref{lem:finrank}:

\begin{proof}
$B=c\, |h><h|$ with $c\geq 0$, i.e., $B\ \varphi = c\, \langle h,\varphi\rangle \
h$ for some $h$. By the min-max principle (see
\cite{ReedSimon4}),
\begin{align}
\tilde{E}_n &= \sup_{\psi_1,\dots,\psi_{n-1}}\,\,
\inf_{\substack{\varphi\perp\psi_1,\dots, \psi_{n-1} \\
||\varphi||=1 }}
\langle \varphi,A \varphi\rangle + c\,| \langle \varphi,h\rangle|^2 \notag \\
&\leq \sup_{\psi_1,\dots ,\psi_{n-1}}\,\, \inf_{\substack {\varphi\perp\psi_1,\dots,\psi_{n-1},h \\
||\varphi||=1 }} \langle \varphi,A \varphi\rangle \notag \\
&\leq \sup_{\psi_1,\dots,\psi_{n-1},\psi_n}\,\,
\inf_{\substack {\varphi\perp\psi_1,\dots,\psi_{n-1},\psi_n\\
||\varphi||=1 }} \langle \varphi,A \varphi  \rangle \notag \\
&=E_{n+1}.
\end{align}
\end{proof}

Wegner's estimate is intimately connected with a method
called ``spectral averaging.'' Roughly speaking, spectral averaging
says that taking expectation with respect to random parameters will
make the spectral measure absolutely continuous. Here is a typical
example which comes from the theory of rank one perturbations, see
Simon \cite{Simon4} and references given there.

To formulate this result, we take a bounded operator $h$ on
$\ell^2(\Z^d)$ and set $h_\alpha=h+\,\alpha\,\delta_j$ for any fixed
$j\in\Z^d$. Note that the multiplication operator $\delta_j$ is a
rank one perturbation of $h$. We denote by $m_\alpha$ the
(projection valued) spectral measure of $h_\alpha$ and set
$\mu_\alpha(A)=\langle \delta_j, m_\alpha(A)\;\delta_j\rangle$. We
obtain

\begin{theorem}[Spectral averaging]
\begin{equation}\label{specaver}
    \int d\,\mu_\alpha(E)\;d\alpha = dE,
\end{equation}
i.e.,
\begin{equation*}
    \int \biggl(\int f(E)\,d\,\mu_\alpha(E)\bigg)\;d\alpha = \int\,f(E)\,dE
\end{equation*}
for all integrable $f$.
\end{theorem}

Wegner's estimate follows from the spectral averaging
result (see \cite{Simon4}).

Spectral averaging was introduced in the theory of random operators
by Kotani \cite{Kotani} who used it to prove Anderson localization.
He used random boundary conditions in dimension one, but soon
Kotani's trick was also used to prove localization in higher-dimensional
systems. However, Kotani was not the first to prove a
spectral averaging formula. Such a formula was known in the Russian
literature earlier, e.g., to Javrjan \cite{Ja} who proved it for
``random'' boundary conditions. Spectral averaging also plays  a
prominent role for continuous Schr\"odinger operators; see \cite{CHM}
and references given there.

\subsection{Regularity in the Continuous Case}
To prove a Wegner estimate for the continuous case is
considerably harder than for the discrete case. In fact, only the
alloy-type model and a few other cases can be treated so far (for these cases see
\cite{CHM} and \cite{FHML}). For the alloy-type model one can carry over Wegner's
original proof (\cite{Kirsch2}).

However, the finite rank estimate (Lemma \ref{lem:finrank}) cannot
be transferred directly to the continuum (see \cite{Kirsch6}). Thus
a direct analog of Wegner's approach only gives
\begin{eqnarray}\label{Wegner21}
\mathbb{E}(N_\Lambda(E+\epsilon)-N_\Lambda(E-\epsilon)) \leq C
|\Lambda|^2 \epsilon \ .
\end{eqnarray}
The estimate (\ref{Wegner21}) obviously gives \emph{no} information
on the regularity of $N$. However, it suffices as input to
multiscale analysis to prove Anderson localization (this was
observed in \cite{MarHol}).

The first to prove a Wegner bound with the ``correct'' volume
dependence were Kotani and Simon in \cite{KoS}. They required the
single site potential $f$ to be the characteristic function of the
unit cube. Later Combes and Hislop \cite{CH1} relaxed this
condition.

Meanwhile, there is a large number of results on ``generalized''
Wegner estimates of the form
\begin{eqnarray}\label{Wegnerka}
\mathbb{E}(N_\Lambda(E+\epsilon)-N_\Lambda(E-\epsilon)) \leq C
|\Lambda|^k \epsilon^\alpha \ .
\end{eqnarray}
For $k=1$, they imply H\"older continuity of the integrated density of
states with H\"older exponent $\alpha$.

We refer to the excellent survey by Veselic \cite{Veselic} on the
subject which describes the development until $2004$. In addition,
we mention the papers
\cite{BCH2,CHK,CHKN,CHN,CHKR,HK,HKV,HS,KSS1,KS1, KS2, KV} for
further reading.

Very recently, Combes, Hislop and Klopp
\cite{CHKnew} published a result which includes (and improves)
virtually all previous results.

\begin{theorem}
Let $q_i$ be independent random variables with a common distribution
$P_0$ of compact support. Let $f$ be a non-negative single site
potential of compact support. Then
\begin{enumerate}
    \item If $P_0$ is H\"older continuous with H\"older exponent
    $\alpha$, then $N$ is H\"older continuous with the same exponent.

    \item If $P_0$ is Lifshitz continuous, then $N$ is
    Lifshitz continuous as well. In this case $N$ has a bounded density.
\end{enumerate}
\end{theorem}

The authors of \cite{CHKnew} actually prove a more general
theorem allowing the random variables to be dependent and including
a magnetic field.

\subsection{Beyond the Density of States: Level Statistics}
One may look at the energy statistics of a disordered
system on a smaller scale than we do for the integrated density of
states. This subject is common in the theory of random matrices
since the days of Wigner and Dyson (see, e.g., \cite{Mehta} or
\cite{Deift}),
but is still in its infancy for random Schr\"odinger operators.

Suppose we have a random Schr\"odinger operator $H$ which we restrict
to a cube $\LL$ by appropriate boundary conditions. We
call the resulting operator $H_{\LL}$. For a fixed cube $\LL$,
the operator $H_{\LL}$ has roughly $|\LL|$
energy levels around an energy $E$. So we may say that the averaged
level spacing near $E$ is $\frac{1}{|\LL|}$. We now look at the
eigenvalues around $E$ under a microscope zooming the averaged level
spacing to $1$. The keyword is ``unfolding of the spectrum.'' By this
we mean we look at the measure
\begin{equation}\label{def:levstat}
    \mu_L([a,b])=\#\bigg\{n; E_n(H_{\LL})\in \biggl[E+\frac{a}{|\LL|},E+\frac{b}{|\LL|}
    \biggr]\biggr\}.
\end{equation}

It is reasonable to ask whether there is a limit of
$\mu_L$ when $L$ goes to infinity. Moreover, if such a limit
exists, what are the properties of the limit measure $\mu$?

Molchanov \cite{Molchanov} proved the existence of this limit for a
one-dimensional model in the continuum. Minami \cite{Minami}
investigated the multidimensional discrete case in the regime of
Anderson localization. Observe that we have localization for all
energies for Molchanov's model.

We discuss Minami's case (Molchanov's case being similar). In the
following we give merely a rough sketch of Minami's result, leaving
out many details---even assumptions and precise statements. A complete
discussion is far beyond the scope of this paper. We urge the reader
to look at the paper \cite{Minami} to get a complete
picture.

Let us suppose we have an integrated density of states $N$ which has
a bounded density $n$, i.e.,
$N(E)=\int_{-\infty}^E\,n(\lambda)\,d\lambda$. For energies near the
bottom of the spectrum, Minami shows that $\mu_L$ is asymptotically a
Poisson measure of intensity $n(E)$. This implies that for $a<b$,
\begin{equation}
\P\big(\mu_L([a,b]=k)\big)\approx\,\frac{n(E)^k(b-a)^k}{k!}\,e^{-n(E)(b-a)} .
\end{equation}
Moreover, the random variables $\mu_L(A_1),\ldots,\mu_L(A_m)$ are
approximately independent.

In an informal (but provable) sense, this means that the eigenvalues
near $E$ look like independent random variables with a uniform
distribution. Especially we have
\begin{align}
  \P\big(\mu_L([a,b]=1)\big) &\approx  n(E)(b-a)\,e^{-n(E)(b-a)}, \\
  \P\big(\mu_L([a,b]=2)\big) &\approx \frac{n(E)^2(b-a)^2}{2}\,e^{-n(E)(b-a)}
\end{align}
and consequently one may prove
$$
\P\bigg(\,\mu_L([-x/2,x/2])\le 2~ \Big|~\mu_L([-x/2,x/2])\le 1\bigg)\approx\,\frac{(n(E)^2 x^2)/2}{n(E)\,x}\sim n(E)\,
x.
$$
This says in a rough way that the \emph{differences} of energy
levels near $E$ have a probability density which is strictly
positive near $0$. In physics terminology, there is no
level repulsion.

One expects that this is not the case in the energy region of
extended states. So far nobody has proven the existence of extended states
for the Anderson model. A fortiori, nobody has proven level
repulsion in this case. However, level repulsion is  well
established for the classical Wigner--Dyson ensembles of random
matrix theory (\cite{Mehta, Deift}).

\end{document}